\documentclass[
submission
]{dmtcs-episciences}

\usepackage[utf8]{inputenc}
\usepackage{subfigure}
\usepackage{amsmath}
\usepackage{amssymb}
\usepackage{amsthm}
\usepackage{hyperref}

\usepackage[round]{natbib}

\newtheorem{definition}{Definition}[section]
\newtheorem{proposition}[definition]{Proposition}
\newtheorem{theorem}[definition]{Theorem}
\newtheorem{lemma}[definition]{Lemma}
\newtheorem{corollary}[definition]{Corollary}
\newtheorem{remark}[definition]{Remark}
\newtheorem{openproblem}[definition]{Open problem}

\newcommand{\example}{\noindent\textbf{Example. }}
\newlength{\taille} \makeatletter
\def\qed{%
  \ifmmode\vrule width .5\baselineskip height 0pt depth .5\baselineskip%
  \else{%
    \unskip\nobreak\hfil%
    \setlength{\taille}{\f@size\p@}%
    \penalty50\hskip1em\null\nobreak\hfil\vrule width .5\taille height
    0pt depth .5\taille
    \parfillskip=0pt\finalhyphendemerits=0\endgraf}%
  \fi} \makeatother

\newlength{\taillepreuve}
\newenvironment{demo}{%
  \setbox123=\hbox{Proof :}%
  \taillepreuve=\wd123%

  \vskip-\lastskip\nobreak\medskip\par\noindent\box123\list{}{\leftmargin
    .5\taillepreuve}\parindent=1em\item} {\qed\endlist\bigskip}

\DeclareMathOperator{\Domain}{Domain}
\DeclareMathOperator{\Lex}{Lex}
\DeclareMathOperator{\ContreLex}{ContreLex}
\DeclareMathOperator{\Hierar}{Hierar}
\DeclareMathOperator{\ContreHierar}{ContreHierar}
\DeclareMathOperator{\Inv}{Inv}
\DeclareMathOperator{\Next}{Next}
\DeclareMathOperator{\Length}{L}
\DeclareMathOperator{\Contre}{Contre}
\DeclareMathOperator{\Prelude}{Prelude}
\DeclareMathOperator{\Postlude}{Postlude}
\DeclareMathOperator{\AntiLex}{AntiLex}
\DeclareMathOperator{\AntiContreLex}{AntiContreLex}
\DeclareMathOperator{\AntiHierar}{AntiHierar}
\DeclareMathOperator{\AntiContreHierar}{AntiContreHierar}
\DeclareMathOperator{\MidLex}{MidLex}
\DeclareMathOperator{\MultiplyCost}{M}

\author{Laurent Lyaudet}
\title[A class of orders with linear? time sorting algorithm]{A class of orders with linear? time sorting algorithm}
\affiliation{\url{https://lyaudet.eu/laurent/}}
\received{2018/12/30}
\revised{}
\accepted{}
\begin{document}
%\publicationdetails{VOL}{2017}{ISS}{NUM}{SUBM}
\maketitle
\begin{abstract}
In this article, we give a precise mathematical meaning to ``linear? time''
that matches experimental behaviour of the algorithm.
The sorting algorithm is not our own, it is a variant of radix sort with counting sort as a subroutine.
The true result of this article is an efficient universality result for lexicographic order,
or more generally for some linear extensions of the partial order ``\begin{math}\Next\end{math}'':
``if current items are equal, compare next items''.
We define new classes of orders: (Finite width) Tree Structured Orders.
We show that an instance of a finite width tree structured order can be converted in linear time and space to an instance of lexicographic order.
The constants implied by the ``nextification'' algorithm are small (around 3 for real world orders).
The class of finite width tree structured orders contains finite orders (\{0, 1\}, int32, int64, ..., float, double, ...),
and orders constructed from them on a tree structure.
In particular, unbounded integers, strings with arbitrary collation, and all orders used for sorting SQL queries are finite width tree structured orders.
\end{abstract}

Current version : 2018/12/30

\section{Introduction}
\label{section:introduction}
The classical complexity result that is taught in a computer science curriculum 
is that sorting can be done in \begin{math}\Theta(n\times \log(n))\end{math}. 
Of course, the word "comparisons" is purposely missing at the end of the previous sentence
(and also the fact that \begin{math}n\end{math} is the number of elements to sort),
but it unfortunately reflects the frequently found informal content of the teachings and beliefs on sorting.
It is a problem both for the \begin{math}\Omega(n\times \log(n))\end{math} and the \begin{math}O(n\times \log(n))\end{math} asymptotic bounds,
as we shall see.
 
First, it is a problem for \begin{math}\Omega(n\times \log(n))\end{math}. 
Indeed, if we consider a finite order of size \begin{math}k\end{math},
the counting sort (see \cite{Seward1954}) solves any instance of \begin{math}n\end{math} elements of this order in time
\begin{math}O(n \times \log(k)) = O(n)\end{math}.

Second, it is also a problem for \begin{math}O(n\times \log(n))\end{math} since this is ``comparison complexity'' instead of time complexity.
It makes the implicit assumption that comparison should have \begin{math}O(1)\end{math} time complexity for practical purposes.
It does not take into account the size of the input ; but it is relatively to input size that we are supposed to measure time complexity.
Let us denote \begin{math}m\end{math} the size of an instance of sorting problem.
There is no polynomial upper bound of \begin{math}m\end{math} given \begin{math}n\end{math}.
This is a situation that is different from what occurs in algorithmic graph theory, for example,
where the complexity of algorithms is given relatively to the number of vertices and edges of the graph.
(The size of a reasonable encoding of a graph is upper bounded by \begin{math}O(n^2 \times \log(n))\end{math},
or \begin{math}O((n+m) \times \log(n))\end{math},
if \begin{math}n\end{math} is the number of vertices,
and \begin{math}m\end{math} is the number of edges.)
However, comparison of long strings with a long common prefix would have \begin{math}\Theta(\frac{m}{n})\end{math} time complexity,
assuming that all strings to be sorted have roughly the same size.
If \begin{math}\MultiplyCost(p)\end{math} denotes the time complexity of multiplication of integers
 of \begin{math}p\end{math} bits,
comparison of two unbounded rational numbers would have \begin{math}\Theta(\MultiplyCost(\frac{m}{n}))\end{math} time complexity (assuming again that all rationals to be sorted have roughly the same size).
(Currently the best asymptotic upper bound known for \begin{math}\MultiplyCost(p)\end{math}
is \begin{math}p \times \log(p) \times 4^{\log^{*}(p)}\end{math} (see \cite{DBLP:journals/corr/abs-1802-07932}).)
These two practical cases would yield \begin{math}\Theta(m \times \log(n))\end{math},
and \begin{math}\Theta(m \times \log(\frac{m}{n}) \times 4^{\log^{*}(\frac{m}{n})} \times \log(n))\end{math}
 time complexity respectively.

Worse yet, we could consider a pathological sort based on an undecidable problem like the halting problem.
An element of the order would be an encoding of a Turing machine with an input for it,
with the knowledge that this Turing machine does halt on this input,
and the order would be based on the number of steps of computation needed for the Turing machine to halt.
(We cannot construct this order but we can define it and thus we can define the associated sorting problem.)
Sorting only two elements would have no computable upper bound on its time complexity,
whilst comparison complexity is 1.
In particular, the general sorting problem is not polynomial time solvable, and not even decidable.
(It would be funny to ask students, computer engineer, and computer scientists
to see how many of them would answer that sorting is polynomial time solvable if asked whether this is true.)

As we can see clearly with these last examples,
\begin{math}\Theta(n\times \log(n))\end{math} comparison complexity result is here for a good reason:
\textbf{We cannot say anything more precise and as general than that.}
But we never explain too much why it is so.

This article tries its best to provide results that are much more precise but at the cost of being slightly less general.
For achieving this goal, we make a paradigm change.
Instead of considering the comparison ``black box'' model,
we open the box and consider ``common'' orders as they are defined and constructed.
It appears that all these common orders are defined using an expression, i.e. a labeled tree,
that these common orders exclude pathological sorts,
and that for them we have \begin{math}\Theta?(m)\end{math} time and space complexity.
We note that many researchers already partially opened the box to consider integers sorting and finite orders sorting,
 and as an aside lexicographic sorting
(see for example \cite{AnderssonHagerupNilssonRaman1998} where they obtained \begin{math}\Theta(n\times \log(\log(n)))\end{math} bounds for finite orders,
 and \begin{math}\Theta(m \times \log(\log(n)))\end{math} bounds for string sorting).
We think the results in our article give an order theory and logical addition to the literature on sorting that complements well these results.

The question mark on asymptotic notation is just a short-hand for some ``conditional complexity''.
In fact every complexity result is conditional ; 
when we write \begin{math}\Theta(n \times \log(n))\end{math} time complexity, we may want to say:
\begin{itemize}
\item assuming Turing machine model, \begin{math}\Theta(n \times \log(n)\ |\ time\ Turing\ machine\ model)\end{math},
\item assuming RAM model, \begin{math}\Theta(n \times \log(n)\ |\ time\ RAM\ model)\end{math}.
\end{itemize}
In this article, we use the question mark to say 
``assuming RAM model where pointers and size integers have constant size and arithmetic on pointers/sizes takes constant time''.
We could have written something like \begin{math}\Theta(m\ |\ time\ RAM,\ pointer/size(size:\ O(1),\ +:\ O(1),\ -:\ O(1),\ \times:\ O(1)))\end{math} instead.
The question mark on asymptotic notations will have this meaning whithin this article,
however we hope it may take other meanings in other articles where it can serve as a handy short-hand.

The justification for this conditional complexity is linked to current 64-bit processors ;
 with 8-bit processors we could not have considered that sorting 256 octets of data is all that we need.
However, we do consider that sorting 16 exbibytes is more than enough for current real world problems.
With 128-bits processors we could roughly sort all the atoms of the earth...

A great success of the black box model is that the algorithms are generic,
 and you just need to provide a callback for the comparison function.
Our results will enable similar reuse of any algorithm suited for doing lexicographic or variants of lexicographic sorting,
and we aim at providing a library (see \url{https://github.com/LLyaudet/TSODLULS}) for generating the code for or dynamically converting the keys of the elements to sort into a simple string.
In order to achieve this goal, we will try to provide a language for describing orders.

Section \ref{section:definitions_and_notations} contains most of the definitions and notations used in this article.
In Section \ref{section:first_results_on_orders}, we build our first results using the definitions, so that the reader gets accustomed to them.
Section \ref{section:universality_of_lexicographic_order} recalls a result of Cantor and proves an efficient version of it for a wide class of orders.
In Section \ref{section:other_results_on_finite_tree_structured_orders}, we prove that finite width tree structured orders are countable orders.
Section \ref{section:orders_that_almost_fit_in_our_framework} looks at the frontiers of our results and open problems.
Finally, Section \ref{section:TSODL} presents an attempt at a tree structured order definition language.

\section{Definitions and notations}
\label{section:definitions_and_notations}

Throughout this article, we use the following definitions and notations.
\begin{math}O\end{math} denotes an order (it is a total/linear order),
in particular \begin{math}O^{0,1}\end{math} denotes the binary order where \begin{math}0 < 1\end{math}.
We denote \begin{math}\Domain(O)\end{math}, the domain of the order \begin{math}O\end{math}
(for example, \begin{math}\Domain(O^{0,1}) = \{0,1\}\end{math}).
\begin{math}\mathcal{O}_i\end{math} denotes a sequence of orders indexed by the ordinal \begin{math}i\end{math},
\begin{math}i = \Length(\mathcal{O}_i)\end{math} is the length of \begin{math}\mathcal{O}_i\end{math},
in particular \begin{math}\mathcal{O}^{0,1}_{\omega} = (O^{0,1})_{\omega}\end{math}
denotes the sequence of binary orders repeated a countable number of times, \begin{math}\omega\end{math} is its length.
It is not necessary to know what an ordinal is to understand 99\% of this article.
Later we define Tree Structured Orders using arbitrary ordinals,
but the algorithmic results are for finite width tree structured orders.
For these later orders we use only the finite ordinals (first, second, third, fourth, etc.),
and the first limit ordinal \begin{math}\omega\end{math} as an upper bound
(in order to express finite but not with a finite upper bound, it can be as big as we want so long it stays finite).
Given two ordinals \begin{math}i < j\end{math}, and a sequence of orders \begin{math}\mathcal{O}_{j}\end{math},
we denote \begin{math}\mathcal{O}_{j}[i] = O_{i}\end{math}, the item of rank \begin{math}i\end{math} in the sequence
(the ranks start at 0).
The reader might know Von Neumann's construction of the ordinals
(an ordinal can be seen as a set that contains exactly all ordinals that are strictly before it, the 0th ordinal is the empty set),
in which case we can consider that \begin{math}i \in j \Leftrightarrow i < j\end{math}.
We also use this notation, for example in \begin{math}\mathcal{O}_{i} = (O_j)_{j \in\ i}\end{math}.
While ordinals are frequently denoted by greek letters, 
we will try to keep using \begin{math}i, j, k, l\end{math} for this purpose,
so that it recalls finite indices to the reader.

We denote \begin{math}\Inv(O)\end{math}, the inverse order of \begin{math}O\end{math};
for example, \begin{math}\Inv(O^{0,1}) = O^{1,0}\end{math} is the order on 0 and 1
where \begin{math}1 < 0\end{math}.
We also denote \begin{math}\Inv(\mathcal{O}_i) = (\Inv(O_j))_{j \in\ i}\end{math},
the sequence of inverse orders of \begin{math}\mathcal{O}_i\end{math};
for example, if \begin{math}\mathcal{O}_3 = (O^{0,1}, O^{0,1}, O^{0,1,2})\end{math},
then \begin{math}\Inv(\mathcal{O}_3) = (O^{1,0}, O^{1,0}, O^{2,1,0})\end{math}
(each order in the sequence is inverted but the ranks of the items are preserved).
Note that \begin{math}\Inv\end{math} is an involution: \begin{math}\Inv(\Inv(O)) = O\end{math}, 
and \begin{math}\Inv(\Inv(\mathcal{O}_i)) = \mathcal{O}_i\end{math}.

In the definitions below, we indicate ``(new term)'' when we used a new word for denoting a mathematical object
that probably exists somewhere else in mathematical literature ;
we indicate ``(new)'' when we defined a new mathematical object
for which we found no anteriority in mathematical literature.
If some reader might provide references for early use of these concepts,
we would be happy to add them here.

\begin{definition}[Prelude sequence (new term)]
Given two ordinals \begin{math}i < j\end{math},
and two sequences of orders \begin{math}\mathcal{O}_{i}, \mathcal{O}_{j}\end{math},
we say that \begin{math}\mathcal{O}_{i}\end{math} is a \emph{prelude (sequence)} of \begin{math}\mathcal{O}_{j}\end{math},
if and only if \begin{math}\mathcal{O}_{i}[k] = \mathcal{O}_{j}[k], \forall k \in i\end{math} (we have \begin{math}k < i < j\end{math}).
We note \begin{math}\Prelude(\mathcal{O}_{j})\end{math} the set of all prelude sequences
 of the order sequence \begin{math}\mathcal{O}_{j}\end{math}.
\end{definition}

For example, \begin{math}(O^{0,1}, O^{0,1})\end{math} is a prelude of \begin{math}(O^{0,1}, O^{0,1}, O^{0,1,2})\end{math};
it is also a prelude of \begin{math}\mathcal{O}^{0,1}_{\omega}\end{math}.

We say that \begin{math}X\end{math} is an element of \begin{math}\mathcal{O}_{i}\end{math} (denoted \begin{math}X \in \mathcal{O}_{i}\end{math}),
when \begin{math}X = (x)_{i}\end{math} is a sequence indexed by \begin{math}i\end{math}
with \begin{math}x_{k} \in \Domain(O_{k}) = \Domain(\mathcal{O}_{i}[k]), \forall k \in i\end{math}.
By convention, there is a unique sequence of orders of length 0 \begin{math}\mathcal{O}_{0}\end{math} ; 
it has only one element denoted \begin{math}\epsilon\end{math} (the empty word).
\begin{math}\mathcal{O}_{0}\end{math} is a prelude sequence of any other order sequence.
We try to avoid confusion by distinguishing \emph{element} \begin{math}X\end{math} of \begin{math}\mathcal{O}_{i}\end{math}
from \emph{item} \begin{math}\mathcal{O}_{j}[i]\end{math} of \begin{math}\mathcal{O}_{i}\end{math};
for example, the ``word'' 002 is an element of \begin{math}\mathcal{O}_3 = (O^{0,1}, O^{0,1}, O^{0,1,2})\end{math},
whilst \begin{math}O^{0,1}\end{math} is the first and second item of \begin{math}\mathcal{O}_3\end{math}.

\begin{definition}[Compatible (sequences) (new)]
Given two sequences of orders \begin{math}\mathcal{O}_{i}, \mathcal{O}_{j}\end{math},
we say that they are \emph{compatible} if they are equal, or one is a prelude sequence of the other.
Let \begin{math}X \in \mathcal{O}_{i}, Y \in \mathcal{O}_{j}\end{math},
we say that \begin{math}X, Y\end{math} are \emph{compatible} if \begin{math}\mathcal{O}_{i}, \mathcal{O}_{j}\end{math} are compatible.
\end{definition}

As we shall see, \emph{compatible} elements are easily converted into \emph{comparable} elements.
Indeed, both ``element-items'' at the same rank in the two elements may be compared,
 since they belong to the domain of the same ``order-item''.
The missing step for obtaining a total order is the object of several definitions below.

\begin{definition}[Question (new term)]
Given two compatible elements \begin{math}X, Y\end{math} of sequences of orders \begin{math}\mathcal{O}_{i}, \mathcal{O}_{j}\end{math},
we say that \begin{math}(k, x_k, y_k)\end{math} is the \emph{question} of \begin{math}X, Y\end{math},
if \begin{math}k\end{math} is the smallest ordinal such that \begin{math}x_k \neq y_k\end{math}.
If \begin{math}X \neq Y\end{math}, and neither \begin{math}X\end{math} is a prefix of \begin{math}Y\end{math},
nor \begin{math}Y\end{math} is a prefix of \begin{math}X\end{math},
 such a \begin{math}k\end{math} exists because ordinals are well-ordered. 
By contrapositive, if no such a \begin{math}k\end{math} exists, then \begin{math}X = Y\end{math},
 or either \begin{math}X\end{math} is a prefix of \begin{math}Y\end{math},
 or \begin{math}Y\end{math} is a prefix of \begin{math}X\end{math}.
 Thus if no such a \begin{math}k\end{math} exists, then \begin{math}X = Y\end{math},
 or \begin{math}\Length(X) \neq \Length(Y)\end{math}.
\end{definition}

\begin{definition}[Order-sequence-operator (new)]
An \emph{order-sequence-operator} is an operator that maps a set of compatible order sequences to an order.
\end{definition}

\begin{lemma}
Let \begin{math}S\end{math} be a set of compatible order sequences, 
then there exists an order sequence \begin{math}\mathcal{O}_{j}\end{math} of which all order sequences of \begin{math}S\end{math} are preludes.
\end{lemma}
\begin{demo}
By definition of compatible sequences, it is clear that for any ordinal \begin{math}i\end{math},
there is at most one sequence of length \begin{math}i\end{math} in \begin{math}S\end{math}.
Since \begin{math}S\end{math} is a set, there exists an ordinal \begin{math}j\end{math} such that all sequences 
of \begin{math}S\end{math} have length less than \begin{math}j\end{math} (because the class of all ordinals is not a set).
Let \begin{math}j\end{math} be the smallest such ordinal (because ordinals are well-ordered).
\begin{itemize}
\item If \begin{math}j\end{math} is a successor ordinal,
 \begin{math}\forall i \in\ j - 1, \forall \mathcal{O}_k \in S, \mathcal{O'}_{k'} \in S\end{math}, 
 with \begin{math}i < k, k' < j\end{math}, we have \begin{math}\mathcal{O}_k[i] = \mathcal{O'}_{k'}[i] = O_i\end{math}.
 Moreover, by minimality of \begin{math}j\end{math}, \begin{math}\forall i \in\ j - 1\end{math},
   \begin{math}\mathcal{O}_k \in S\end{math}, \begin{math}\mathcal{O}_k[i]\end{math}, and \begin{math}O_i\end{math} exist.
 Let us denote \begin{math}\mathcal{O}_j\end{math}, the sequence of all \begin{math}O_i\end{math} padded with \begin{math}O^{0,1}_{j-1}\end{math}.
 It is clear that \begin{math}\mathcal{O}_j\end{math} satisfies that all order sequences of \begin{math}S\end{math} are preludes of it.
\item If \begin{math}j\end{math} is a limit ordinal,
 \begin{math}\forall i \in\ j, \forall \mathcal{O}_k \in S, \mathcal{O'}_{k'} \in S\end{math}, 
 with \begin{math}i < k, k' < j\end{math}, we have \begin{math}\mathcal{O}_k[i] = \mathcal{O'}_{k'}[i] = O_i\end{math}.
 Moreover, by minimality of \begin{math}j\end{math}, \begin{math}\forall i \in\ j\end{math},
   \begin{math}\mathcal{O}_k \in S\end{math}, \begin{math}\mathcal{O}_k[i]\end{math}, and \begin{math}O_i\end{math} exist.
 Let us denote \begin{math}\mathcal{O}_j\end{math}, the sequence of all \begin{math}O_i\end{math}.
 It is clear again that \begin{math}\mathcal{O}_j\end{math} satisfies that all order sequences of \begin{math}S\end{math} are preludes of it.
\end{itemize}
\end{demo}

Using these first lemma and definitions, we shall now define five order-sequence-operators.
The notations may be surprising because instead of considering arbitrary sets of compatible order sequences,
we will consider all preludes of some order sequence for four of them. 
(The orders constructed with arbitrary sets would be sub-orders of the orders that we will define now.)
The only step toward arbitrary sets of compatible order sequences will be a possible restriction on the minimum length of the sequences.
This step will be justified by optimizations and other results later in this article.

We start by a ``partial-order-sequence-operator''; its result is a partial order instead of a total order.
\begin{definition}[Next partial order]
Given two ordinals \begin{math}i < j\end{math},
and a sequence of orders \begin{math}\mathcal{O}_{j} = (O_k)_{k \in\ j}\end{math},
the \emph{next partial order} denoted \begin{math}\Next(i, j, \mathcal{O}_{j})\end{math}
is a partial order defined on the set of all elements of the preludes of \begin{math}\mathcal{O}_{j}\end{math},
such that these preludes have length at least \begin{math}i\end{math}.
This is the partial order satisfying \begin{math}\forall X, Y \in \Domain(\Next(i, j, \mathcal{O}_{j}))\end{math}
(the lengths \begin{math}\Length(X), \Length(Y)\end{math} of \begin{math}X\end{math} and \begin{math}Y\end{math}
are such that \begin{math}i \leq \Length(X), \Length(Y) < j\end{math}),
\begin{itemize}
\item if \begin{math}X, Y\end{math} have a question \begin{math}(k, x_k, y_k)\end{math}, then:
\begin{itemize}
  \item if \begin{math}x_k < y_k\end{math} in \begin{math}O_{k}\end{math}, then \begin{math}X < Y\end{math},
  \item if \begin{math}x_k > y_k\end{math} in \begin{math}O_{k}\end{math}, then \begin{math}X > Y\end{math},
\end{itemize}
\item if they don't have a question, then elements are not ordered.
\end{itemize}
\end{definition}

The partial order \begin{math}\Next(i, j, \mathcal{O}_{j})\end{math}
corresponds to the simple idea for comparing sequences
``if current items are equal, compare next items''.

\begin{definition}[Lexicographic order]
Given two ordinals \begin{math}i < j\end{math},
and a sequence of orders \begin{math}\mathcal{O}_{j} = (O_k)_{k \in\ j}\end{math},
the \emph{lexicographic order} denoted \begin{math}\Lex(i, j, \mathcal{O}_{j})\end{math}
is an order defined on the set of all elements of the preludes of \begin{math}\mathcal{O}_{j}\end{math},
such that these preludes have length at least \begin{math}i\end{math}.
This is the order satisfying \begin{math}\forall X, Y \in \Domain(\Lex(i, j, \mathcal{O}_{j}))\end{math},
\begin{itemize}
\item if \begin{math}X, Y\end{math} have a question \begin{math}(k, x_k, y_k)\end{math}, then:
\begin{itemize}
  \item if \begin{math}x_k < y_k\end{math} in \begin{math}O_{k}\end{math}, then \begin{math}X < Y\end{math},
  \item if \begin{math}x_k > y_k\end{math} in \begin{math}O_{k}\end{math}, then \begin{math}X > Y\end{math},
\end{itemize}
\item if they don't have a question, then either \begin{math}X = Y\end{math}, or \begin{math}\Length(X) \neq \Length(Y)\end{math}: 
\begin{itemize}
  \item if \begin{math}\Length(X) < \Length(Y)\end{math}, then \begin{math}X < Y\end{math},
  \item if \begin{math}\Length(X) > \Length(Y)\end{math}, then \begin{math}X > Y\end{math}.
\end{itemize}
\end{itemize}
\end{definition}

\example \begin{math}\Lex(0, \omega, \mathcal{O}^{0,1}_{\omega})\end{math} is the lexicographic order on all finite binary strings, including the empty word.

Lexicographic order may be found under many names (see wikipedia):
\begin{itemize}
  \item lexicographical order,
  \item lexical order,
  \item dictionary order,
  \item alphabet order (-- the term ``alphabet order'' should be reserved for the order on the letters, even if the lexicographic order is based on it recursively),
  \item lexicographic(al) product.
\end{itemize}

\begin{definition}[Contre-Lexicographic order (new term)]
Given two ordinals \begin{math}i < j\end{math},
and a sequence of orders \begin{math}\mathcal{O}_{j} = (O_k)_{k \in\ j}\end{math},
the \emph{contre-lexicographic order} denoted \begin{math}\ContreLex(i, j, \mathcal{O}_{j})\end{math}
is an order defined on the set of all elements of the preludes of \begin{math}\mathcal{O}_{j}\end{math},
such that these preludes have length at least \begin{math}i\end{math}.
This is the order satisfying \begin{math}\forall X, Y \in \Domain(\ContreLex(i, j, \mathcal{O}_{j}))\end{math},
\begin{itemize}
\item if \begin{math}X, Y\end{math} have a question \begin{math}(k, x_k, y_k)\end{math}, then:
\begin{itemize}
  \item if \begin{math}x_k < y_k\end{math} in \begin{math}O_{k}\end{math}, then \begin{math}X < Y\end{math},
  \item if \begin{math}x_k > y_k\end{math} in \begin{math}O_{k}\end{math}, then \begin{math}X > Y\end{math},
\end{itemize}
\item if they don't have a question, then either \begin{math}X = Y\end{math}, or \begin{math}\Length(X) \neq \Length(Y)\end{math}:
\begin{itemize}
  \item if \begin{math}\Length(X) < \Length(Y)\end{math}, then \begin{math}X > Y\end{math},
  \item if \begin{math}\Length(X) > \Length(Y)\end{math}, then \begin{math}X < Y\end{math}.
\end{itemize}
\end{itemize}
\end{definition}

Lexicographic and contre-lexicographic orders are the two simplest linear extensions of the partial order 
\begin{math}\Next(i, j, \mathcal{O}_{j})\end{math}.
With lexicographic order, if no difference is found in the items of the sequence, then the shortest sequence comes first.
With contre-lexicographic order, if no difference is found in the items of the sequence, then the shortest sequence comes last.

\begin{definition}[Hierarchic order]
Given two ordinals \begin{math}i < j\end{math},
and a sequence of orders \begin{math}\mathcal{O}_{j} = (O_k)_{k \in\ j}\end{math},
the \emph{hierarchic order} denoted \begin{math}\Hierar(i, j, \mathcal{O}_{j})\end{math}
is an order defined on the set of all elements of the preludes of \begin{math}\mathcal{O}_{j}\end{math},
such that these preludes have length at least \begin{math}i\end{math}.
This is the order satisfying \begin{math}\forall X, Y \in \Domain(\Hierar(i, j, \mathcal{O}_{j}))\end{math},
\begin{itemize}
  \item if \begin{math}\Length(X) < \Length(Y)\end{math}, then \begin{math}X < Y\end{math},
  \item if \begin{math}\Length(X) > \Length(Y)\end{math}, then \begin{math}X > Y\end{math},
  \item otherwise \begin{math}X = Y\end{math}, or \begin{math}X, Y\end{math} have a question \begin{math}(k, x_k, y_k)\end{math}, then:
\begin{itemize}
  \item if \begin{math}x_k < y_k\end{math} in \begin{math}O_{k}\end{math}, then \begin{math}X < Y\end{math},
  \item if \begin{math}x_k > y_k\end{math} in \begin{math}O_{k}\end{math}, then \begin{math}X > Y\end{math}.
\end{itemize}
\end{itemize}
\end{definition}

\example \begin{math}\Hierar(0, \omega, \mathcal{O}^{0,1}_{\omega})\end{math} 
(\begin{math}\Hierar(0, \omega, \mathcal{O}^{0,1,2,3,4,5,6,7,8,9}_{\omega})\end{math}) 
is the hierarchic order on all finite binary strings (finite decimal strings respectively).
If you keep only the string \begin{math}0\end{math} and all strings with a leading non-zero digit,
then it is the order on base 2 integers writing (base 10 integers writing respectively) that matches the order on the integers.
Thus it is the order we use without naming it when we compare two written integers.
The funny fact is that we need to compare the lengths of these strings first.
These lengths are also integers but much smaller (logarithmic)
 and for ``real life'' integers our brain just reads the lengths as they are written in unary.
We do not think that our brain applies recursive thinking to compare the length,
however we can do so consciously for really long integers by counting the number of digits.

Hierarchic(al) order has also many names (some of them are listed on wikipedia shortlex).
One can found it under the names :
\begin{itemize}
  \item radix order,
  \item military order 
  (it is the same idea that hierarchical order ; it could be also political order ; guess who deserves an empty word ? ;) ),
  \item pseudo-lexicographic order,
  \item short(-)lex(icographic) order,
  \item length-lexicographic order,
  \item genealogical order,
  \item lexicographic (-- We found a book where the author used only one order throughout and he kept the only name familiar to him).
\end{itemize}
Words in hierarchic order may be found in dictionaries suited for crosswords and other word games.

\begin{definition}[Contre-Hierarchic order (new term)]
Given two ordinals \begin{math}i < j\end{math},
and a sequence of orders \begin{math}\mathcal{O}_{j} = (O_k)_{k \in\ j}\end{math},
the \emph{contre-hierarchic order} denoted \begin{math}\ContreHierar(i, j, \mathcal{O}_{j})\end{math}
is an order defined on the set of all elements of the preludes of \begin{math}\mathcal{O}_{j}\end{math},
such that these preludes have length at least \begin{math}i\end{math}.
This is the order satisfying \begin{math}\forall X, Y \in \Domain(\ContreHierar(i, j, \mathcal{O}_{j}))\end{math},
\begin{itemize}
  \item if \begin{math}\Length(X) < \Length(Y)\end{math}, then \begin{math}X > Y\end{math},
  \item if \begin{math}\Length(X) > \Length(Y)\end{math}, then \begin{math}X < Y\end{math},
  \item otherwise \begin{math}X = Y\end{math}, or \begin{math}X, Y\end{math} have a question \begin{math}(k, x_k, y_k)\end{math}, then:
\begin{itemize}
  \item if \begin{math}x_k < y_k\end{math} in \begin{math}O_{k}\end{math}, then \begin{math}X < Y\end{math},
  \item if \begin{math}x_k > y_k\end{math} in \begin{math}O_{k}\end{math}, then \begin{math}X > Y\end{math}.
\end{itemize}
\end{itemize}
\end{definition}

As with \begin{math}\ContreLex\end{math}, for \begin{math}\ContreHierar\end{math}, shortest elements comes last.

\begin{definition}[Generalized sum of orders (new term)]
Given a ``master order'' \begin{math}O_m\end{math},
and an injective mapping \begin{math}f\end{math} that associates an order \begin{math}f(x)\end{math}
to each element \begin{math}x\end{math} of \begin{math}O_m\end{math}, the \emph{generalized sum} denoted \begin{math}\sum_{x \in\ O_m} f(x)\end{math}
is an order defined on the disjoint union of all elements of the \begin{math}f(x)\end{math} orders,
 or with redundant information on a subset of the cartesian product between \begin{math}O_m\end{math} 
 and the disjoint union of all elements of the \begin{math}f(x)\end{math} orders.
Thus an element of this order may be ordered as a couple \begin{math}X = (x,y \in\ f(x))\end{math}.
For two elements \begin{math}X = (x,y \in\ f(x))\end{math} and \begin{math}X' = (x',y' \in\ f(x'))\end{math},
we have \begin{math}X < X'\end{math} if and only if \begin{math}x < x'\end{math},
 or \begin{math}x = x'\end{math} and \begin{math}y < y'\end{math} in the order \begin{math}f(x)\end{math}.
\end{definition}
We call this operator a generalized sum because when \begin{math}O_m = O^{0,1}\end{math},
\begin{math}\sum_{x \in\ O_m} f(x)\end{math} is the order sum \begin{math}f(0) + f(1)\end{math}.
The order sum has also many names: linear sum, ordinal sum, series composition, star product.
If \begin{math}f(x) = copy(O), \forall x \in O_m\end{math}, we obtain the order product (linear product, ordinal product) \begin{math}O \times O_m\end{math}
(this product is \begin{math}O\end{math} repeated in \begin{math}O_m\end{math}, each element of \begin{math}O_m\end{math} is replaced by a copy of \begin{math}O\end{math}).

The generalized sum does not look like the other \emph{order-sequence-operators}.
However, applying Zorn's axiom to the domain of \begin{math}O_m\end{math}, we can well-order the elements \begin{math}x\end{math} of \begin{math}O_m\end{math},
and we can replace the mapping \begin{math}f\end{math} by a sequence \begin{math}\mathcal{O}_j\end{math} indexed by an ordinal,
 and thus we can denote \begin{math}\sum_{x \in\ O_m} f(x) = \sum(O_m, \mathcal{O}_j)\end{math}
 where it is implied that \begin{math}O_m\end{math} and \begin{math}\mathcal{O}_j\end{math} are correlated by an implicit well-ordering of \begin{math}O_m\end{math}.
Since there is no efficient way to apply Zorn's axiom, we shall put severe restrictions on \begin{math}O_m\end{math} for efficient results later on.

Note that, whilst hierarchic order enables to order unsigned integers of arbitrary precision,
 it is easy with generalized sum to obtain signed integers of arbitrary precision.

Indeed it is a suborder of \begin{math}\sum(O^{0,1}, ( \Inv(\Hierar(1, \omega, \mathcal{O}^{0,1})), \Hierar(1, \omega, \mathcal{O}^{0,1}) ))\end{math}.

\section{First results on orders}
\label{section:first_results_on_orders}

Let us compare the orders we can construct so far with the building blocks of the previous section.

\hspace{-2cm}\begin{tabular}{|c|c|c|c|}
\hline

\begin{math}\Lex(0, 3, \mathcal{O}^{0,1}_{3})\end{math}
  & \begin{math}\ContreLex(0, 3, \mathcal{O}^{0,1}_{3})\end{math}
  & \begin{math}\Lex(0, 3, \Inv(\mathcal{O}^{0,1}_{3}))\end{math}
  & \begin{math}\ContreLex(0, 3, \Inv(\mathcal{O}^{0,1}_{3}))\end{math}\\

= & = & = & =\\

\begin{math}\Inv(\ContreLex(0, 3, \Inv(\mathcal{O}^{0,1}_{3})))\end{math}
  & \begin{math}\Inv(\Lex(0, 3, \Inv(\mathcal{O}^{0,1}_{3})))\end{math}
  & \begin{math}\Inv(\ContreLex(0, 3, \mathcal{O}^{0,1}_{3}))\end{math}
  & \begin{math}\Inv(\Lex(0, 3, \mathcal{O}^{0,1}_{3}))\end{math}\\

\hline

\begin{math}\epsilon\end{math} & \begin{math}00\end{math} & \begin{math}\epsilon\end{math} & \begin{math}11\end{math}\\
\begin{math}0\end{math} & \begin{math}01\end{math} & \begin{math}1\end{math} & \begin{math}10\end{math}\\
\begin{math}00\end{math} & \begin{math}0\end{math} & \begin{math}11\end{math} & \begin{math}1\end{math}\\
\begin{math}01\end{math} & \begin{math}10\end{math} & \begin{math}10\end{math} & \begin{math}01\end{math}\\
\begin{math}1\end{math} & \begin{math}11\end{math} & \begin{math}0\end{math} & \begin{math}00\end{math}\\
\begin{math}10\end{math} & \begin{math}1\end{math} & \begin{math}01\end{math} & \begin{math}0\end{math}\\
\begin{math}11\end{math} & \begin{math}\epsilon\end{math} & \begin{math}00\end{math} & \begin{math}\epsilon\end{math}\\ 
\hline
\end{tabular}

\hspace{-2cm}\begin{tabular}{|c|c|c|c|}
\hline

\begin{math}\Hierar(0, 3, \mathcal{O}^{0,1}_{3})\end{math}
  & \begin{math}\ContreHierar(0, 3, \mathcal{O}^{0,1}_{3})\end{math}
  & \begin{math}\Hierar(0, 3, \Inv(\mathcal{O}^{0,1}_{3}))\end{math}
  & \begin{math}\ContreHierar(0, 3, \Inv(\mathcal{O}^{0,1}_{3}))\end{math}\\

= & = & = & =\\

\begin{math}\Inv(\ContreHierar(0, 3, \Inv(\mathcal{O}^{0,1}_{3})))\end{math}
  & \begin{math}\Inv(\Hierar(0, 3, \Inv(\mathcal{O}^{0,1}_{3})))\end{math}
  & \begin{math}\Inv(\ContreHierar(0, 3, \mathcal{O}^{0,1}_{3}))\end{math}
  & \begin{math}\Inv(\Hierar(0, 3, \mathcal{O}^{0,1}_{3}))\end{math}\\

\hline

\begin{math}\epsilon\end{math} & \begin{math}00\end{math} & \begin{math}\epsilon\end{math} & \begin{math}11\end{math}\\
\begin{math}0\end{math} & \begin{math}01\end{math} & \begin{math}1\end{math} & \begin{math}10\end{math}\\
\begin{math}1\end{math} & \begin{math}10\end{math} & \begin{math}0\end{math} & \begin{math}01\end{math}\\
\begin{math}00\end{math} & \begin{math}11\end{math} & \begin{math}11\end{math} & \begin{math}00\end{math}\\
\begin{math}01\end{math} & \begin{math}0\end{math} & \begin{math}10\end{math} & \begin{math}1\end{math}\\
\begin{math}10\end{math} & \begin{math}1\end{math} & \begin{math}01\end{math} & \begin{math}0\end{math}\\
\begin{math}11\end{math} & \begin{math}\epsilon\end{math} & \begin{math}00\end{math} & \begin{math}\epsilon\end{math}\\ 
\hline
\end{tabular}

The equalities in the header of the two tables are not dependent of the ordinals or the binary order,
and our first results will be to prove them in their full generality.

\begin{remark}
Given two ordinals \begin{math}i < j\end{math},
 a sequence of orders \begin{math}\mathcal{O}_{j}\end{math},
 and two order-sequence-operators \begin{math}\mu, \nu\end{math} in \begin{math}\{\Lex, \ContreLex, \Hierar, \ContreHierar \}\end{math},
 \begin{math}\Domain(\mu(i, j, \mathcal{O}_{j})) = \Domain(\mu(i, j, \Inv(\mathcal{O}_{j}))) = \Domain(\nu(i, j, \mathcal{O}_{j})) = \Domain(\nu(i, j, \Inv(\mathcal{O}_{j})))\end{math}.
\end{remark}

\begin{proposition}\label{ContreLexFromLexInv}
Given two ordinals \begin{math}i < j\end{math},
and a sequence of orders \begin{math}\mathcal{O}_{j}\end{math},
 \begin{math}\ContreLex(i, j, \mathcal{O}_{j}) = \Inv(\Lex(i, j, \Inv(\mathcal{O}_{j})))\end{math}.
\end{proposition}
\begin{demo}
Let \begin{math} X, Y \in \Domain(\ContreLex(i, j, \mathcal{O}_{j})) = \Domain(\Inv(\Lex(i, j, \Inv(\mathcal{O}_{j}))))\end{math}
No order-sequence-operator changes equality. Hence we need only to prove that
\begin{math}X < Y\end{math} for \begin{math}\ContreLex(i, j, \mathcal{O}_{j})\end{math} if and only if 
\begin{math}X < Y\end{math} for \begin{math}\Inv(\Lex(i, j, \Inv(\mathcal{O}_{j})))\end{math}.

Suppose \begin{math}X < Y\end{math} for \begin{math}\ContreLex(i, j, \mathcal{O}_{j})\end{math}.
\begin{itemize}
\item Either \begin{math}X, Y\end{math} have a question \begin{math}(k, x_k, y_k)\end{math},
  in which case \begin{math}x_k < y_k\end{math} in \begin{math}O_k\end{math}. 
  Hence \begin{math}x_k > y_k\end{math} in \begin{math}\Inv(O_k)\end{math},
  and \begin{math}X > Y\end{math} in \begin{math}\Lex(i, j, \Inv(\mathcal{O}_{j}))\end{math}.
  Thus \begin{math}X < Y\end{math} in \begin{math}\Inv(\Lex(i, j, \Inv(\mathcal{O}_{j})))\end{math}.
\item Or \begin{math}Y\end{math} is a prefix of \begin{math}X\end{math}.
  Hence \begin{math}X > Y\end{math} in \begin{math}\Lex(i, j, \Inv(\mathcal{O}_{j}))\end{math}.
  Thus \begin{math}X < Y\end{math} in \begin{math}\Inv(\Lex(i, j, \Inv(\mathcal{O}_{j})))\end{math}.
\end{itemize}

Suppose \begin{math}X < Y\end{math} for \begin{math}\Inv(\Lex(i, j, \Inv(\mathcal{O}_{j})))\end{math}.
\begin{itemize}
\item Either \begin{math}X, Y\end{math} have a question \begin{math}(k, x_k, y_k)\end{math},
  in which case \begin{math}x_k < y_k\end{math} in \begin{math}\Inv(\Inv(O_k)) = O_k\end{math}. 
  Thus \begin{math}X < Y\end{math} in \begin{math}\ContreLex(i, j, \mathcal{O}_{j})\end{math}.
\item Or \begin{math}Y\end{math} is a prefix of \begin{math}X\end{math}.
  Thus \begin{math}X < Y\end{math} in \begin{math}\ContreLex(i, j, \mathcal{O}_{j})\end{math}.
\item Or \begin{math}X\end{math} is a prefix of \begin{math}Y\end{math}.
  Hence \begin{math}X < Y\end{math} in \begin{math}\Lex(i, j, \Inv(\mathcal{O}_{j}))\end{math}.
  Thus \begin{math}X > Y\end{math} in \begin{math}\Inv(\Lex(i, j, \Inv(\mathcal{O}_{j})))\end{math}.
  A contradiction so this case: \begin{math}X\end{math} is a prefix of \begin{math}Y\end{math}, was not possible.
\end{itemize}
\end{demo}

\begin{corollary}\label{FromLexInv}
Given two ordinals \begin{math}i < j\end{math},
and a sequence of orders \begin{math}\mathcal{O}_{j}\end{math},
 \begin{math}\Lex(i, j, \mathcal{O}_{j}) = \Inv(\ContreLex(i, j, \Inv(\mathcal{O}_{j})))\end{math},
 \begin{math}\Lex(i, j, \Inv(\mathcal{O}_{j})) = \Inv(\ContreLex(i, j, \mathcal{O}_{j}))\end{math},
and  \begin{math}\Inv(\Lex(i, j, \mathcal{O}_{j})) = \ContreLex(i, j, \Inv(\mathcal{O}_{j}))\end{math}.
\end{corollary}

\begin{proposition}\label{ContreHierarFromHierarInv}
Given two ordinals \begin{math}i < j\end{math},
and a sequence of orders \begin{math}\mathcal{O}_{j}\end{math},
 \begin{math}\ContreHierar(i, j, \mathcal{O}_{j}) = \Inv(\Hierar(i, j, \Inv(\mathcal{O}_{j})))\end{math}.
\end{proposition}
\begin{demo}
Let \begin{math} X, Y \in \Domain(\ContreHierar(i, j, \mathcal{O}_{j})) = \Domain(\Inv(\Hierar(i, j, \Inv(\mathcal{O}_{j}))))\end{math}
No order-sequence-operator changes equality. Hence we need only to prove that
\begin{math}X < Y\end{math} for \begin{math}\ContreHierar(i, j, \mathcal{O}_{j})\end{math} if and only if 
\begin{math}X < Y\end{math} for \begin{math}\Inv(\Hierar(i, j, \Inv(\mathcal{O}_{j})))\end{math}.

Suppose \begin{math}X < Y\end{math} for \begin{math}\ContreHierar(i, j, \mathcal{O}_{j})\end{math}.
\begin{itemize}
\item Either \begin{math}\Length(Y) < \Length(X)\end{math}.
  Hence \begin{math}X > Y\end{math} in \begin{math}\Hierar(i, j, \Inv(\mathcal{O}_{j}))\end{math}.
  Thus \begin{math}X < Y\end{math} in \begin{math}\Inv(\Hierar(i, j, \Inv(\mathcal{O}_{j})))\end{math}.
\item Or \begin{math}X, Y\end{math} have a question \begin{math}(k, x_k, y_k)\end{math},
  in which case \begin{math}x_k < y_k\end{math} in \begin{math}O_k\end{math}. 
  Hence \begin{math}x_k > y_k\end{math} in \begin{math}\Inv(O_k)\end{math},
  and \begin{math}X > Y\end{math} in \begin{math}\Hierar(i, j, \Inv(\mathcal{O}_{j}))\end{math}.
  Thus \begin{math}X < Y\end{math} in \begin{math}\Inv(\Hierar(i, j, \Inv(\mathcal{O}_{j})))\end{math}.
\end{itemize}

Suppose \begin{math}X < Y\end{math} for \begin{math}\Inv(\Hierar(i, j, \Inv(\mathcal{O}_{j})))\end{math}.
\begin{itemize}
\item Either \begin{math}\Length(Y) < \Length(X)\end{math}.
  Thus \begin{math}X < Y\end{math} in \begin{math}\ContreHierar(i, j, \mathcal{O}_{j})\end{math}.
\item Or \begin{math}\Length(X) < \Length(Y)\end{math}.
  Hence \begin{math}X < Y\end{math} in \begin{math}\Hierar(i, j, \Inv(\mathcal{O}_{j}))\end{math}.
  Thus \begin{math}X > Y\end{math} in \begin{math}\Inv(\Hierar(i, j, \Inv(\mathcal{O}_{j})))\end{math}.
  A contradiction so this case: \begin{math}\Length(X) < \Length(Y)\end{math}, was not possible.
\item Or \begin{math}X, Y\end{math} have a question \begin{math}(k, x_k, y_k)\end{math},
  in which case \begin{math}x_k < y_k\end{math} in \begin{math}\Inv(\Inv(O_k)) = O_k\end{math}. 
  Thus \begin{math}X < Y\end{math} in \begin{math}\ContreHierar(i, j, \mathcal{O}_{j})\end{math}.
\end{itemize}
\end{demo}

\begin{corollary}\label{FromHierarInv}
Given two ordinals \begin{math}i < j\end{math},
and a sequence of orders \begin{math}\mathcal{O}_{j}\end{math},
 \begin{math}\Hierar(i, j, \mathcal{O}_{j}) = \Inv(\ContreHierar(i, j, \Inv(\mathcal{O}_{j})))\end{math},
 \begin{math}\Hierar(i, j, \Inv(\mathcal{O}_{j})) = \Inv(\ContreHierar(i, j, \mathcal{O}_{j}))\end{math},
and  \begin{math}\Inv(\Hierar(i, j, \mathcal{O}_{j})) = \ContreHierar(i, j, \Inv(\mathcal{O}_{j}))\end{math}.
\end{corollary}

In light of these results, it appears that we could define another involution:
\begin{math}\Contre\end{math} that operates on order-sequence-operators,
\begin{math}\Contre(operator(.)) = \Inv(operator(\Inv(.)))\end{math}. 
Clearly \begin{math}\Contre(\Contre(operator(.))) = operator(.)\end{math}.
Thus we have \begin{math}\Lex = \Contre(\ContreLex)\end{math}, \begin{math}\ContreLex = \Contre(\Lex)\end{math},
 \begin{math}\Hierar = \Contre(\ContreHierar)\end{math}, \begin{math}\ContreHierar = \Contre(\Hierar)\end{math}.

In later results, both operators \begin{math}\Lex\end{math} and \begin{math}\ContreLex\end{math} will be needed
 for labelling the internal nodes of the trees defining orders, if we want to exclude \begin{math}\Inv\end{math}.
We could have also given names to \begin{math}\Lex(i, j, \Inv(\mathcal{O}_{j}))\end{math},
and \begin{math}\Inv(\Lex(i, j, \mathcal{O}_{j}))\end{math} and use these two operators instead.

We shall now prove that \begin{math}\sum = \Contre(\sum)\end{math},
 the generalized sum of orders is its own contre-order.

\begin{proposition}\label{ContreSum}
Given a ``master order'' \begin{math}O_m\end{math},
and a corresponding sequence of orders \begin{math}\mathcal{O}_{j}\end{math},

\begin{math}\Inv(\sum(O_m, \mathcal{O}_j)) = \sum(\Inv(O_m), \Inv(\mathcal{O}_j))\end{math}.
\end{proposition}
\begin{demo}
Clearly \begin{math}\Domain(\Inv(\sum(O_m, \mathcal{O}_j))) = \Domain(\sum(\Inv(O_m), \Inv(\mathcal{O}_j)))\end{math}
 since it is the disjoint union of the orders \begin{math}\mathcal{O}_j\end{math}, or \begin{math}\Inv(\mathcal{O}_j)\end{math} equivalently.
Let \begin{math} X = (x, y), X' = (x', y')\end{math} be two elements of this domain.
\begin{itemize}
\item If \begin{math}x < x'\end{math} in \begin{math}O_m\end{math}, then:
  \begin{itemize}
  \item \begin{math}X < X'\end{math} in \begin{math}\sum(O_m, \mathcal{O}_j)\end{math},
   and \begin{math}X' < X\end{math} in \begin{math}\Inv(\sum(O_m, \mathcal{O}_j))\end{math},
  \item \begin{math}x' < x\end{math} in \begin{math}\Inv(O_m)\end{math},
   and again \begin{math}X' < X\end{math} but in \begin{math}\sum(\Inv(O_m), \Inv(\mathcal{O}_j))\end{math}.
  \end{itemize}

\item If \begin{math}x' < x\end{math} in \begin{math}O_m\end{math}, then, by symmetry with the first case,
  \begin{math}X < X'\end{math} both in \begin{math}\Inv(\sum(O_m, \mathcal{O}_j))\end{math}
   and in \begin{math}\sum(\Inv(O_m), \Inv(\mathcal{O}_j))\end{math}.

\item If \begin{math}x = x'\end{math} in \begin{math}O_m\end{math},
  then there exists some \begin{math}O_i \in \mathcal{O}_j\end{math} of which \begin{math}y\end{math} and \begin{math}y'\end{math} are elements.
  \begin{itemize}
  \item If \begin{math}y < y'\end{math} in \begin{math}O_i\end{math}, \begin{math}X < X'\end{math} in  \begin{math}\sum(O_m, \mathcal{O}_j)\end{math},
   and \begin{math}X' < X\end{math} in \begin{math}\Inv(\sum(O_m, \mathcal{O}_j))\end{math}.
   And we have also that \begin{math}y' < y\end{math} in \begin{math}\Inv(O_i)\end{math},
   thus \begin{math}X' < X\end{math} but in \begin{math}\sum(\Inv(O_m), \Inv(\mathcal{O}_j))\end{math}.
  \item If \begin{math}y' < y\end{math}, by symmetry with the preceding subcase
    \begin{math}X < X'\end{math} both in \begin{math}\Inv(\sum(O_m, \mathcal{O}_j))\end{math}
     and in \begin{math}\sum(\Inv(O_m), \Inv(\mathcal{O}_j))\end{math}..
  \end{itemize}
\end{itemize}
\end{demo}

\section{Universality of lexicographic order}
\label{section:universality_of_lexicographic_order}

\subsection{Theoretical universality}
\label{subsection:theoretical_universality}

\begin{definition}[Universal order]
We say that an order \begin{math}O\end{math} is \emph{universal} for a class of orders \begin{math}\mathcal{A}\end{math},
 if for any order \begin{math}O' \in \mathcal{A}\end{math},
 there exists an embedding of \begin{math}O'\end{math} in \begin{math}O\end{math}.
An order embedding is an injective mapping such that order is preserved.
\end{definition}

This is a classical result by \cite{Cantor1895}
that the order on the rationals \rationals~ (between 0 and 1) is universal for countable orders.
It is easy to see that the same applies for the lexicographic order on binary words.
We adapt the proof to lexicographic order and give it here for the sake of completeness.

\begin{theorem}[Cantor 1895]
\begin{math}\Lex(1, \omega, \mathcal{O}^{0,1}_{\omega})\end{math} is universal for countable orders.
\end{theorem}
\begin{demo}
Let \begin{math}O\end{math} be a countable order and \begin{math}(x_i)_{i \in\ \naturals}\end{math} be an enumeration of its elements.
We associate to each element inductively a rational number between \begin{math}0\end{math} and \begin{math}1\end{math}
 for which the denominator is a power of two.
Since these numbers are strictly between \begin{math}0\end{math} and \begin{math}1\end{math},
there is a bijection between these numbers and their decimal part,
which can be written as a word on the digits 0,1.
Since the denominator is a power of two, the length is finite.
Since the numbers are strictly positive, the length is at least one.
The last digit is a 1.

The first element is associated to \begin{math}\frac{1}{2}\end{math}, i.e. the word ``1''.
For each element after the first, assume by induction, that all previous elements were embedded 
in \begin{math}\Lex(1, \omega, \mathcal{O}^{0,1}_{\omega})\end{math}.

If the current element is the smallest of all elements considered so far,
then assign to it the number \begin{math}\frac{n}{2}\end{math},
where \begin{math}n\end{math} is the rational number associated to the smallest of all previous elements.
Since \begin{math}n\end{math} has a denominator that is a power of two, the same is true for \begin{math}\frac{n}{2}\end{math}.

If the current element is the biggest of all elements considered so far,
then assign to it the number \begin{math}\frac{1+n}{2}\end{math},
where \begin{math}n\end{math} is the rational number associated to the biggest of all previous elements.
Since \begin{math}n\end{math} has a denominator that is a power of two, the same is true for \begin{math}\frac{1+n}{2}\end{math}.

If the current element is between two elements considered so far,
then assign to it the number \begin{math}\frac{n + m}{2}\end{math},
where \begin{math}n, m\end{math} are the rational numbers associated to these two elements.
Since \begin{math}n, m\end{math} have a denominator that is a power of two, the same is true for \begin{math}\frac{n + m}{2}\end{math}.
\end{demo}

Even the pathological example of the introduction (sorting Turing machines by halting number) fits in this theorem.
However, it is not efficient.
Nobody will provide us with an enumeration of some arbitrarily chosen countable order
and an oracle to know where to insert the new element.
Our goal in this article is to prove that for a wide class of orders, the lexicographic order
(or any one of its three variants) is efficiently universal.

Before we do so, we remark that hierarchic order of a sequence of finite orders cannot be universal.
Indeed, there is a finite number of elements of length at most \begin{math}k\end{math}.
Thus, there is no infinite descending chain in this order.
Since it is a total order (or linear order), there is no infinite anti-chain (all chains have size 1),
hierachic order is well-ordered ; so it cannot embeds arbitrary countable orders.
The same reasoning applies to contre-hierachic order since it doesn't contain infinite ascending chain.

The generalized sum of order cannot produce infinite orders from finite orders so we will not try to use it for universality results.

\subsection{Efficient universality}
\label{subsection:efficient_universality}

Let us now define the class of orders for which we will have efficient embeddings in lexicographic order over binary strings.
\begin{definition}[Tree structured orders]
\emph{Tree structured orders} are constructed recursively as follow:
\begin{itemize}
\item Finite orders are tree structured orders.
\item Given a tree structured order \begin{math}O\end{math}, \begin{math}\Inv(O)\end{math} is a tree structured order.
\item Given two ordinals \begin{math}i < j\end{math}, and a sequence of tree structured orders \begin{math}\mathcal{O}_{j}\end{math},
  \begin{math}\Lex(i, j, \mathcal{O}_{j})\end{math} is a tree structured order.
\item Given two ordinals \begin{math}i < j\end{math}, and a sequence of tree structured orders \begin{math}\mathcal{O}_{j}\end{math},
  \begin{math}\Hierar(i, j, \mathcal{O}_{j})\end{math} is a tree structured order.
\item Given a master tree structured order \begin{math}O_m\end{math},
  and a corresponding sequence of tree structured orders \begin{math}\mathcal{O}_{j}\end{math},
  \begin{math}\sum(O_m, \mathcal{O}_j)\end{math} is a tree structured order.
\end{itemize}
We also have that (by propositions \ref{ContreLexFromLexInv}, and \ref{ContreHierarFromHierarInv}):
\begin{itemize}
\item Given two ordinals \begin{math}i < j\end{math}, and a sequence of tree structured orders \begin{math}\mathcal{O}_{j}\end{math},
  \begin{math}\ContreLex(i, j, \mathcal{O}_{j})\end{math} is a tree structured order.
\item Given two ordinals \begin{math}i < j\end{math}, and a sequence of tree structured orders \begin{math}\mathcal{O}_{j}\end{math},
  \begin{math}\ContreHierar(i, j, \mathcal{O}_{j})\end{math} is a tree structured order.
\end{itemize} 
\end{definition}

Thus any tree structured order is defined by a labelled tree with possibly infinite depth and infinite width.
Note that any path from the root of the tree to a leaf is finite, but because we have infinite sequences of orders,
the depth of the tree may be infinite.
The leaves of this labelled tree are finite orders.
The internal nodes are labelled with two ordinals and an order-sequence-operator.
In order to have efficient universality, we have to restrict ourself to more finitely described orders.
We cannot consider arbitrary sequences of orders.
Instead we will consider ultimately periodic sequences ; the period will be written between [ ],
 for example \begin{math}\mathcal{O}_{\omega} = (O^{0,1}, [O^{0,1}, O^{1,0}])_{\omega}\end{math} is the sequence
where the first item and all items of even rank are the binary order,
whilst all items of odd rank greater than one are the inverse of the binary order (we considered that rank started at 1 for this explanation).

When the sequence is periodic with a period of length 1, like \begin{math}\mathcal{O}^{0,1}_{\omega} = ([O^{0,1}])_{\omega}\end{math},
we say that the sequence is uniform.

\begin{definition}[Finite width tree structured orders]
\emph{Finite width tree structured orders} are constructed recursively as follow:
\begin{itemize}
\item Finite orders are finite width tree structured orders.
\item Given a finite width tree structured order \begin{math}O\end{math}, \begin{math}\Inv(O)\end{math} is a finite width tree structured order.
\item Given two ordinals \begin{math}i < j \leq \omega\end{math},
  and an ultimately periodic sequence of finite width tree structured orders \begin{math}\mathcal{O}_{j}\end{math},
  \begin{math}\Lex(i, j, \mathcal{O}_{j})\end{math} is a finite width tree structured order.
\item Given two ordinals \begin{math}i < j \leq \omega\end{math},
  and an ultimately periodic sequence of finite width tree structured orders \begin{math}\mathcal{O}_{j}\end{math},
  \begin{math}\Hierar(i, j, \mathcal{O}_{j})\end{math} is a finite width tree structured order.
\item Given a master finite order \begin{math}O_m\end{math},
  and a corresponding sequence of finite width tree structured orders \begin{math}\mathcal{O}_{j}\end{math},
  \begin{math}\sum(O_m, \mathcal{O}_j)\end{math} is a finite width tree structured order.
\end{itemize}
We also have that (by propositions \ref{ContreLexFromLexInv}, and \ref{ContreHierarFromHierarInv}):
\begin{itemize}
\item Given two ordinals \begin{math}i < j \leq \omega\end{math},
  and an ultimately periodic sequence of finite width tree structured orders \begin{math}\mathcal{O}_{j}\end{math},
  \begin{math}\ContreLex(i, j, \mathcal{O}_{j})\end{math} is a finite width tree structured order.
\item Given two ordinals \begin{math}i < j \leq \omega\end{math},
  and an ultimately periodic sequence of finite width tree structured orders \begin{math}\mathcal{O}_{j}\end{math},
  \begin{math}\ContreHierar(i, j, \mathcal{O}_{j})\end{math} is a finite width tree structured order.
\end{itemize} 
\end{definition}

With these restrictions and using [ ] notation on sub-expressions or sub-trees,
 finite width tree structured orders can be represented by a finite expression and/or a finite labelled tree.

We used \begin{math}\Inv, \Lex, \Hierar, \sum\end{math} to define (finite width) tree structured order,
 however, by propositions and corollaries \ref{ContreLexFromLexInv}, \ref{FromLexInv},
 \ref{ContreHierarFromHierarInv}, \ref{FromHierarInv}, \ref{ContreSum}
 we could have chosen the following sets of operators instead: 
\begin{itemize}
\item \begin{math}\Inv, \ContreLex, \Hierar, \sum\end{math}, 
\item \begin{math}\Inv, \Lex, \ContreHierar, \sum\end{math},
\item \begin{math}\Inv, \ContreLex, \ContreHierar, \sum\end{math},
\item \begin{math}\Lex, \ContreLex, \Hierar, \ContreHierar, \sum\end{math}.
\end{itemize}
It is trivial to see except maybe for the last one.
For \begin{math}\Lex, \ContreLex, \Hierar, \ContreHierar, \sum\end{math} we have to remark that
 we can ``push'' all \begin{math}\Inv\end{math} nodes from the root of the tree to its leaves.
Indeed we can exchange an internal node labelled with an order-sequence-operator together with its parent \begin{math}\Inv\end{math} node,
 for an internal node with the ``contre'' order-sequence-operator and children \begin{math}\Inv\end{math} nodes between the original children 
 of the order-sequence-operator node.
Since the depth of the tree is finite, we end up with \begin{math}\Inv\end{math} nodes only above leaves,
 where they can disappear since the inverse of a finite order is a finite order.

We continue with a remark where almost every order-sequence-operator considered so far is ``universal''.
\begin{remark}[Fundamental remark on orders]\label{fundamental_remark}
For fixed length sequences of orders, there is no difference at all between (contre-)lexicographic, and (contre-)hierarchic order.
For any ordinal \begin{math}i\end{math}, and any order sequence \begin{math}\mathcal{O}_{i+1}\end{math},
\begin{math}\Lex(i, i + 1, \mathcal{O}_{i+1}) = \ContreLex(i, i + 1, \mathcal{O}_{i+1}) = \Hierar(i, i + 1, \mathcal{O}_{i+1}) = \ContreHierar(i, i + 1, \mathcal{O}_{i+1}) = \Next(i, i + 1, \mathcal{O}_{i+1})\end{math}.
\end{remark}

Whilst we never saw this remark explicitly stated, it is used very frequently,
 and is probably the source of the confusion between lexicographic and hierarchic order.
It renders useless to name the operator, for example in SQL queries:
\textbf{SELECT User.* FROM User \emph{ORDER BY User.login, User.birthdate, User.firstname, User.lastname}} ;
a few commas are sufficient.
It explains that radix sort that was created for fixed length integers should in reality be considered a string sorting algorithm,
for strings of arbitrary size.

What we obtain with fixed length, we will try to achieve for arbitrary length.
It will require that the length itself, or more precisely the end of the string, is explicitely encoded in the string
so that everything is sorted \emph{before} we actually reach the end of the string.
It means that we will not only prove that lexicographic order is efficiently universal,
but also that contre-lexicographic order and ``next'' partial order are efficiently universal.
A ``mirror work'' could be done with \begin{math}\Inv(\Next), \Inv(\Lex), \Inv(\ContreLex)\end{math}.

From the fundamental remark on orders, we can deduce our encoding of finite orders.
We enumerate the elements in the finite order from the smallest to the greatest and we use fixed length integer to do so.
Thus finite orders are embedded in \begin{math}\Next(i, i+1, \mathcal{O}^{0,1}_{i+1})\end{math}, for some integer \begin{math}i\end{math}.
(We must remark that we assume that the used finite orders are not ``obfuscated''.
Indeed, for example, the order \begin{math}O^{0,1}\end{math} could be obfuscated behind the halting problem as in the pathological example of the introduction.
The elements to be sorted would be Turing machine encodings with an input for them and the key of the sort is 0 if the Turing machine halts on its input, 1 otherwise.
We assume that the keys for sorting finite orders are given explicitly or can be computed in constant time from the element to be sorted.
Any efficient result has to make this assumption.)

What would happen if we had an \begin{math}\Inv\end{math} node?
If the order below has been ``binarily nextified'', then all we need to do is switch 0s and 1s 
so that the new strings are ordered correctly by ``(contre-)lex/next''.

Let us look at other internal nodes.

If we have a \begin{math}\Lex\end{math} node, and everything below has been binarily nextified,
 then we almost only have to contatenate the strings.
Almost because we want to have an indicator that the obtained string is finished.
Maybe we could decrement some counter placed at the end of string.
But for that this counter must be initialized to some value that is greater than the number of \begin{math}\Lex\end{math} node from a leaf to the tree.

The \begin{math}\ContreLex\end{math} node will also require a counter that we could increment.

\begin{definition}[TSO-padding (new)]
Given a finite width tree structured order with tree/expression \begin{math}T\end{math},
 let \begin{math}l\end{math} be an integer such that \begin{math}2^l\end{math} is greater than
  the maximum number of \begin{math}\Lex\end{math} nodes on a path from a leaf of \begin{math}T\end{math} to its root;
 let \begin{math}c\end{math} be an integer such that \begin{math}2^c\end{math} is greater than
  the maximum number of \begin{math}\ContreLex\end{math} nodes on a path from a leaf of \begin{math}T\end{math} to its root.
A \begin{math}l\end{math}-lex-padding is a set of \begin{math}l\end{math} contiguous bits that will serve as a counter for \begin{math}\Lex\end{math} nodes.
A \begin{math}c\end{math}-contrelex-padding is a set of \begin{math}c\end{math} contiguous bits that will serve as a counter for \begin{math}\ContreLex\end{math} nodes.
A \begin{math}(l,c)\end{math}-\emph{TSO-padding} is a \begin{math}l\end{math}-lex-padding
and a \begin{math}c\end{math}-contrelex-padding that are contiguous.
\end{definition}

For the rest of this subsection, it will be clearer to the reader if she assumes that the tree \begin{math}T\end{math} has depth less than 15,
which is more than enough for ``common'' orders.
So that we can have TSO-paddings that are one octet where the first four bits are the lex-padding and the last four bits are the contrelex-padding.
Let us assume in the following that a byte = an octet is the smallest allocatable space.
The default value for TSO-paddings is \begin{math}(2^l - 1, 0)\end{math} (assume it is \begin{math}(15, 0)\end{math}).

We can encode a finite order by adding after each byte of the original encoding a TSO-padding byte.
However, how do we encode an empty string at the level of a \begin{math}\Lex\end{math} node?
If we encode an empty string with a single TSO-padding byte,
 we have a shift between this TSO-padding byte at rank 0 that should be compared with TSO-padding bytes at odd rank.
There are two simple possibilities:
\begin{itemize}
\item We can padd along the repeated pattern (padding0, data1, padding2).
  Thus the empty string in a \begin{math}\Lex\end{math} node (or \begin{math}\ContreLex\end{math} node)
  would be matched to a pattern with padding0 = (0, 0), data1 = 0, padding2 = (15,0).
  It would be slightly unefficient since padding0 bytes would serve only to deal with empty string,
  but still it would be linear space with a constant of 3.
\item We can consider that there is no empty string and that all strings should be ``null started''.
  Hence we use the repeated pattern (data0, padding1).
  All non-empty strings start with a pattern data0 = 0, padding1 = (15,0)
  The empty string is coded by the pattern data0 = 0, padding1 = (0,0).
  The problem is: What happens if an empty string is the last item to be concatenated in an above
  \begin{math}\Lex\end{math} node (or \begin{math}\ContreLex\end{math} node)?
  If we decrement the final lex-padding, since it is 0, we will have a problem.
  Another attempt where the empty string would be coded by data0 = anything and padding1 = (d,0), where d is the depth of the node,
  would fail to distinguish an empty string at depth d from an empty string at depth d+1,
  for which the lex-counter was decremented because it is the only item at depth d.
  We need to encode the depth of empty strings but we cannot decrement/increment this information.
\end{itemize}
These two simple possibilities show something important: making the padding right may be technical and open to traps,
 so in what follows we will use the repeated pattern (padding0, data1, padding2) for its robustness and simplicity.
However for some finite width tree structured order, more efficient paddings can be possible.
(When there is no path between the root and a leaf with more than one \begin{math}\Lex\end{math} or \begin{math}\ContreLex\end{math} node,
or when all such nodes have strictly positive minimum length of preludes, we can use the pattern (data0, padding1).
When we have only a finite order, we don't need padding.
When all the leaves correspond to finite orders encoded using exactly \begin{math}k\end{math} bytes,
 then we can use the repeated pattern (padding0, data1, padding2) where data1 is \begin{math}k\end{math} bytes, instead of only one byte.
Etc.)
It is one of our goal (out of scope for this article) to code a library that does the right choices for padding
 according to each particular tree/expression given to describe an order.

\begin{definition}[TSO-encoding (new)]
We say that an element of an order is \emph{TSO-encoded}
if it is represented by a binary string that contains TSO-padding along the repeated pattern (padding0, data1, padding2),
 so that binary \begin{math}\Next\end{math} partial order applied to it and to another \emph{TSO-encoded} string for the same order 
yields the same result than the original order.
We may say that this binary string is the \emph{TSO-encoding} of the element.
\end{definition}

Given a finite order encoded by integers with \begin{math}k\end{math} bytes,
we obtain in linear time and space a TSO-encoded string with \begin{math}k\end{math} times the pattern 
(padding0 = (15, 0), data1, padding2 = (15,0)), and we decrement the final padding2 to (14,0).

Assume we obtained a TSO-encoding for all children suborders.
Assume also for the length of this paragraph that we did remove the \begin{math}\Inv\end{math} nodes
 as we explained after defining finite width tree structured orders.
\begin{itemize}
\item If we have a \begin{math}\Lex\end{math} node above, we can concatenate in linear time and space the TSO-encodings of the suborders
and decrement the lex-value of the final padding2.
If the length of the prelude for this \begin{math}\Lex\end{math} node is zero, 
we can create in constant time and space a pattern padding0 = (d, 0), data1 = 0, padding2 = (15, 0),
where d is the depth of the \begin{math}\Lex\end{math} node in the tree (it should be 0 if it is the root node).
Let \begin{math}X, Y\end{math} be elements of the order obtained at this node,
and \begin{math}E(X), E(Y)\end{math} be their obtained encoding.
If \begin{math}X, Y\end{math} have a question \begin{math}(k, x_k, y_k)\end{math}, then all subelements before the question are equal
 and thus \begin{math}E(X), E(Y)\end{math} starts with a common prefix (that may be empty)
 before we compare the patterns encoding \begin{math}x_k, y_k\end{math}.
But we assumed these patterns were TSO-encodings, hence comparing these patterns yields the same result as comparing \begin{math}x_k, y_k\end{math}.
If they don't have a question and assuming they are different, we can suppose, without loss of generality,
that \begin{math}X\end{math} is a prefix of \begin{math}Y\end{math};
 if \begin{math}X\end{math} is the empty string,
 then the starting padding0 of \begin{math}E(X)\end{math} has lex-value lower than the corresponding padding0 of \begin{math}E(Y)\end{math}
 (even if \begin{math}E(Y)\end{math} started with the encoding of an empty string from a subnode);
 if \begin{math}X\end{math} is not the empty string, then the final padding2 of \begin{math}E(X)\end{math} has lex-value lower than the corresponding padding2 of \begin{math}E(Y)\end{math} and bytes before it are equal in \begin{math}E(X)\end{math} and \begin{math}E(Y)\end{math};
 hence comparing these patterns yields \begin{math}X < Y\end{math}.

\item If we have a \begin{math}\ContreLex\end{math} node above, we can concatenate in linear time and space the TSO-encodings of the suborders
and increment the contrelex-value of the final padding2.
If the length of the prelude for this \begin{math}\ContreLex\end{math} node is zero, 
we can create in constant time and space a pattern padding0 = (15, 15 - d), data1 = 0, padding2 = (15, 0),
where d is the depth of the \begin{math}\Lex\end{math} node in the tree (it should be 0 if it is the root node).
Let \begin{math}X, Y\end{math} be elements of the order obtained at this node,
and \begin{math}E(X), E(Y)\end{math} be their obtained encoding.
If \begin{math}X, Y\end{math} have a question \begin{math}(k, x_k, y_k)\end{math}, then all subelements before the question are equal
 and thus \begin{math}E(X), E(Y)\end{math} starts with a common prefix (that may be empty)
 before we compare the patterns encoding \begin{math}x_k, y_k\end{math}.
But we assumed these patterns were TSO-encodings, hence comparing these patterns yields the same result as comparing \begin{math}x_k, y_k\end{math}.
If they don't have a question and assuming they are different, we can suppose, without loss of generality,
that \begin{math}X\end{math} is a prefix of \begin{math}Y\end{math};
 if \begin{math}X\end{math} is the empty string,
 then the starting padding0 of \begin{math}E(X)\end{math} has lex-value and contrelex-value greater than the corresponding padding0 of \begin{math}E(Y)\end{math};
 if \begin{math}X\end{math} is not the empty string, then the final padding2 of \begin{math}E(X)\end{math} has contrelex-value greater than the corresponding padding2 of \begin{math}E(Y)\end{math} and bytes before it are equal in \begin{math}E(X)\end{math} and \begin{math}E(Y)\end{math};
 hence comparing these patterns yields \begin{math}X < Y\end{math}.

\item If we have a \begin{math}\Hierar\end{math} node above, 
we can encode the number of subelements as follow: 
we encode in unary the number of bytes needed to express the byte length of the number of subelements,
then we concatenate the bytes encoding the byte length of the number of subelements,
 and finally we concatenate the binary encoding of the number of subelements.
(Let us give an example, if the number of subelements is \begin{math}2^{400}\end{math},
 the byte length of the number of subelements is \begin{math}\lceil 401 / 8 \rceil = 51\end{math},
 the number of bytes needed to encode this length is 1.
So we encode 10000000 in unary for the first byte, then 00110011 = 51 for the following byte,
 then the 51 bytes representing \begin{math}2^{400}\end{math}.
(It is important that the unary encoding is terminated by a 0.
Thus, if the number of bytes needed to encode the byte length of the number of subelements was 8,
then the unary encoding would be 11111111 00000000.
It would require that the number of subelements is greater than
\begin{math}2^{8\times (2^{56}-1)}-1\end{math}, which is way more than the number of atoms in the universe.))
The needed number of bytes is in \begin{math}O(\log(n) + \log(\log(n))\end{math} where \begin{math}n\end{math} is the number of subelements.

All these number of subelements bytes are padded along the common pattern (padding0, data1, padding2)
 as we would for finite orders.
Then we concatenate the TSO-encodings of the subelements.
It is trivial to see that if the numbers of subelements are different then the corresponding patterns will be ordered accordingly,
otherwise the concatenations of TSO-encodings of the subelements will be ordered correctly.
Note that the empty string with 0 subelement is already handled by the number of elements encoding (10000000 10000000 00000000),
but we could simplify with a single null byte.
Note also that we do not have to increment/decrement anything.

\item If we have a \begin{math}\ContreHierar\end{math} node above,
 similarly to \begin{math}\Hierar\end{math} node, we encode the number of subelements but we then switch 0s and 1s in the obtained encoding,
 before padding.

\item If we have a \begin{math}\sum\end{math} node above,
 we can treat this node as a \begin{math}\Lex,\Next\end{math} node with a sequence of orders of length 2.
 Thus we just need to concatenate the TSO-encoding of the element of \begin{math}O_m\end{math}, 
 with the TSO-encoding of the element of \begin{math}f(x)\end{math}.
 It is not a problem that we have TSO-encodings for several distinct orders \begin{math}f(x), \forall x \in O_m\end{math}.
 Indeed such encodings will ever be compared if they correspond to the same order,
 since the encoding of the element of \begin{math}O_m\end{math} is compared first.
\end{itemize}

\begin{math}\Inv\end{math} node had the simplest way to convert ``binarily nextified'' encodings that were not yet TSO-encodings.
However switching 0s and 1s will be a problem with padding2 TSO-paddings since we may decrement the lex-part and/or increment the contre-lex part later on.
What can we do about it? 
Lex-part and contre-lex part can be in this order or in the other order without altering the result.
Let's just do that.
If the depth of the finite width tree structured order is less than 15, then, after switching 0s and 1s,
we exchange the four bits of lex-padding and the four bits of contre-lex padding of the last padding2.
Its simple arithmetic to see that, after that, we can increment/decrement until we reach the root node.
Note that for performance, we would probably remove the \begin{math}\Inv\end{math} nodes,
 since switching 0s and 1s on all the string constructed so far would be costly.
A little rewriting on the tree/expression would be more efficient if the number of elements to sort is enough.

\begin{theorem}
\begin{math}\Lex(1, \omega, \mathcal{O}^{0,1}_{\omega})\end{math},
 \begin{math}\ContreLex(1, \omega, \mathcal{O}^{0,1}_{\omega})\end{math},
 and \begin{math}\Next(1, \omega, \mathcal{O}^{0,1}_{\omega})\end{math} are efficiently universal
 for finite width tree structured orders
 (and for orders that can be efficiently embedded in finite width tree structured orders).
 Given a finite width tree structured order \begin{math}O\end{math} and its tree/expression \begin{math}T\end{math},
 there is a linear? time and space algorithm that takes as input \begin{math}T\end{math} 
 and an array \begin{math}A\end{math} of elements of \begin{math}O\end{math} 
 and converts \begin{math}A\end{math} into an array of cells where each cell contains 
 a pointer to the original element of \begin{math}O\end{math}
 and another pointer to a binary string suited for ``(contre-)lex/next'' sorting.
\end{theorem}
\begin{demo}
The content of this subsection so far was the proof of this theorem.
\end{demo}

\begin{corollary}
The order on the rationals \rationals\ between 0 and 1 is efficiently universal
for finite width tree structured orders.
 Given a finite width tree structured order \begin{math}O\end{math} and its tree/expression \begin{math}T\end{math},
 there is a linear? time and space algorithm that takes as input \begin{math}T\end{math}
 and an array \begin{math}A\end{math} of elements of \begin{math}O\end{math}
 and converts \begin{math}A\end{math} into an array of cells where each cell contains
 a pointer to the original element of \begin{math}O\end{math}
 and a binary string coding the decimal part of a rational number between 0 and 1.
\end{corollary}

\subsection{Other nodes for defining orders}
\label{subsection:other_nodes_for_defining_orders}

We finish this section with syntactic sugar, i.e. new nodes for defining orders.
These nodes doesn't add new orders to the class of tree structured orders.

\begin{definition}[Postlude sequence (new term)]
Given two \emph{finite} ordinals \begin{math}i < j\end{math},
and two sequences of orders \begin{math}\mathcal{O}_{i}, \mathcal{O}_{j}\end{math},
we say that \begin{math}\mathcal{O}_{i}\end{math} is a \emph{postlude (sequence)} of \begin{math}\mathcal{O}_{j}\end{math},
if and only if \begin{math}\mathcal{O}_{i}[k] = \mathcal{O}_{j}[k + (j - i)], \forall k \in i\end{math} (we have \begin{math}k < i < j\end{math}).
We note \begin{math}\Postlude(\mathcal{O}_{j})\end{math} the set of all postlude sequences
 of the order sequence \begin{math}\mathcal{O}_{j}\end{math}.
\end{definition}

\begin{definition}[Anti-lexicographic order]
Given two \emph{finite} ordinals \begin{math}i < j\end{math},
and an uniform sequence of orders \begin{math}\mathcal{O}_{j} = ([O])_{j}\end{math} (\begin{math}O\end{math} repeated \begin{math}j\end{math} times),
the \emph{anti-lexicographic order} denoted \begin{math}\AntiLex(i, j, \mathcal{O}_{j})\end{math}
is an order defined on the set of all elements of the postludes of \begin{math}\mathcal{O}_{j}\end{math},
such that these postludes have length at least \begin{math}i\end{math}.
This is the order satisfying \begin{math}\forall X, Y \in \Domain(\AntiLex(i, j, \mathcal{O}_{j}))\end{math}
(the lengths \begin{math}\Length(X), \Length(Y)\end{math} of \begin{math}X\end{math} and \begin{math}Y\end{math}
are such that \begin{math}i \leq \Length(X), \Length(Y) < j\end{math}),
\begin{itemize}
\item if there exists a finite ordinal \begin{math} 0 < k \leq \min(\Length(X), \Length(Y))\end{math}
 such that \begin{math}x_{\Length(X) - k} \neq y_{\Length(Y) - k}\end{math}, then:
\begin{itemize}
  \item if \begin{math}x_{\Length(X) - k} < y_{\Length(Y) - k}\end{math} in \begin{math}O\end{math}, then \begin{math}X < Y\end{math},
  \item if \begin{math}x_{\Length(X) - k} > y_{\Length(Y) - k}\end{math} in \begin{math}O\end{math}, then \begin{math}X > Y\end{math},
\end{itemize}
\item otherwise no such \begin{math}k\end{math} exists, then either \begin{math}X = Y\end{math}, or \begin{math}\Length(X) \neq \Length(Y)\end{math}: 
\begin{itemize}
  \item if \begin{math}\Length(X) < \Length(Y)\end{math}, then \begin{math}X < Y\end{math},
  \item if \begin{math}\Length(X) > \Length(Y)\end{math}, then \begin{math}X > Y\end{math}.
\end{itemize}
\end{itemize}
\end{definition}

Anti-lexicographic order is just lexicographic order made backward.
Thus it cannot work for arbitrary ordinals ; it works only for finite ordinals.
Assuming that the software objects we are applying anti-lexicographic order (or some other order-sequence-operator)
are themselves sorted in order to yield the expected result, 
it may be useful to have \begin{math}\AntiLex\end{math} nodes in order to avoid reversing an array of objects.
One can treat these nodes with a for loop concatenating the suborders TSO-encoded strings backward
(each substring is written forward but the set of substrings is treated backward, then the final lex-padding is decremented).

Similarly, one can define anti-contre-lexicographic order, anti-hierarchic order and anti-contre-hierarchic order for finite sequences of orders.
Here are the definitions for the sake of completeness.

\begin{definition}[Anti-contre-lexicographic order (new-term)]
Given two \emph{finite} ordinals \begin{math}i < j\end{math},
and an uniform sequence of orders \begin{math}\mathcal{O}_{j} = ([O])_{j}\end{math} (\begin{math}O\end{math} repeated \begin{math}j\end{math} times),
the \emph{anti-contre-lexicographic order} denoted \begin{math}\AntiContreLex(i, j, \mathcal{O}_{j})\end{math}
is an order defined on the set of all elements of the postludes of \begin{math}\mathcal{O}_{j}\end{math},
such that these postludes have length at least \begin{math}i\end{math}.
This is the order satisfying \begin{math}\forall X, Y \in \Domain(\AntiContreLex(i, j, \mathcal{O}_{j}))\end{math},
\begin{itemize}
\item if there exists a finite ordinal \begin{math} 0 < k \leq \min(\Length(X), \Length(Y))\end{math}
 such that \begin{math}x_{\Length(X) - k} \neq y_{\Length(Y) - k}\end{math}, then:
\begin{itemize}
  \item if \begin{math}x_{\Length(X) - k} < y_{\Length(Y) - k}\end{math} in \begin{math}O\end{math}, then \begin{math}X < Y\end{math},
  \item if \begin{math}x_{\Length(X) - k} > y_{\Length(Y) - k}\end{math} in \begin{math}O\end{math}, then \begin{math}X > Y\end{math},
\end{itemize}
\item otherwise no such \begin{math}k\end{math} exists, then either \begin{math}X = Y\end{math}, or \begin{math}\Length(X) \neq \Length(Y)\end{math}: 
\begin{itemize}
  \item if \begin{math}\Length(X) < \Length(Y)\end{math}, then \begin{math}X > Y\end{math},
  \item if \begin{math}\Length(X) > \Length(Y)\end{math}, then \begin{math}X < Y\end{math}.
\end{itemize}
\end{itemize}
\end{definition}

\begin{definition}[Anti-hierarchic order (new term)]
Given two \emph{finite} ordinals \begin{math}i < j\end{math},
and an uniform sequence of orders \begin{math}\mathcal{O}_{j} = ([O])_{j}\end{math} (\begin{math}O\end{math} repeated \begin{math}j\end{math} times),
the \emph{anti-hierarchic order} denoted \begin{math}\AntiHierar(i, j, \mathcal{O}_{j})\end{math}
is an order defined on the set of all elements of the postludes of \begin{math}\mathcal{O}_{j}\end{math},
such that these postludes have length at least \begin{math}i\end{math}.
This is the order satisfying \begin{math}\forall X, Y \in \Domain(\AntiHierar(i, j, \mathcal{O}_{j}))\end{math},
\begin{itemize}
  \item if \begin{math}\Length(X) < \Length(Y)\end{math}, then \begin{math}X < Y\end{math},
  \item if \begin{math}\Length(X) > \Length(Y)\end{math}, then \begin{math}X > Y\end{math}.
  \item otherwise \begin{math}X = Y\end{math}, or there exists a finite ordinal \begin{math} 0 < k \leq \Length(X) = \Length(Y)\end{math}
   such that \begin{math}x_{\Length(X) - k} \neq y_{\Length(Y) - k}\end{math}, then:
  \begin{itemize}
    \item if \begin{math}x_{\Length(X) - k} < y_{\Length(Y) - k}\end{math} in \begin{math}O\end{math}, then \begin{math}X < Y\end{math},
    \item if \begin{math}x_{\Length(X) - k} > y_{\Length(Y) - k}\end{math} in \begin{math}O\end{math}, then \begin{math}X > Y\end{math}.
  \end{itemize}
\end{itemize}
\end{definition}

\begin{definition}[Anti-contre-hierarchic order (new term)]
Given two \emph{finite} ordinals \begin{math}i < j\end{math},
and an uniform sequence of orders \begin{math}\mathcal{O}_{j} = ([O])_{j}\end{math} (\begin{math}O\end{math} repeated \begin{math}j\end{math} times),
the \emph{anti-contre-hierarchic order} denoted \begin{math}\AntiContreHierar(i, j, \mathcal{O}_{j})\end{math}
is an order defined on the set of all elements of the postludes of \begin{math}\mathcal{O}_{j}\end{math},
such that these postludes have length at least \begin{math}i\end{math}.
This is the order satisfying \begin{math}\forall X, Y \in \Domain(\AntiContreHierar(i, j, \mathcal{O}_{j}))\end{math},
\begin{itemize}
  \item if \begin{math}\Length(X) < \Length(Y)\end{math}, then \begin{math}X > Y\end{math},
  \item if \begin{math}\Length(X) > \Length(Y)\end{math}, then \begin{math}X < Y\end{math}.
  \item otherwise \begin{math}X = Y\end{math}, or there exists a finite ordinal \begin{math} 0 < k \leq \Length(X) = \Length(Y)\end{math}
   such that \begin{math}x_{\Length(X) - k} \neq y_{\Length(Y) - k}\end{math}, then:
  \begin{itemize}
    \item if \begin{math}x_{\Length(X) - k} < y_{\Length(Y) - k}\end{math} in \begin{math}O\end{math}, then \begin{math}X < Y\end{math},
    \item if \begin{math}x_{\Length(X) - k} > y_{\Length(Y) - k}\end{math} in \begin{math}O\end{math}, then \begin{math}X > Y\end{math}.
  \end{itemize}
\end{itemize}
\end{definition}

\section{Other results on finite width tree structured orders}
\label{section:other_results_on_finite_tree_structured_orders}

\begin{theorem}
For any finite width tree structured order, its domain is countable.
\end{theorem}
\begin{demo}
We can deduce this theorem from the fact that any element of this order can be associated (efficiently) to a finite binary string.
However, we give here a proof that does not need this previous result.
The proof follows the definition and uses the fact that a countable union of countable sets is countable.
\begin{itemize}
\item Finite orders are countable.
\item Given a countable order \begin{math}O\end{math}, \begin{math}\Inv(O)\end{math} is a countable order.
\item Given two ordinals \begin{math}i < j \leq \omega\end{math}, and an ultimately periodic sequence of countable orders \begin{math}\mathcal{O}_{j}\end{math},
  \begin{math}\Prelude(\mathcal{O}_{j})\end{math} is countable and each prelude has finite length,
  hence the set of elements of the prelude are a finite union of countable sets, so are countable.
  Thus the union of elements of the sequences in \begin{math}\Prelude(\mathcal{O}_{j}))\end{math}
  are a countable union of countable sets, hence this union is countable.
  \begin{math}\Lex(i, j, \mathcal{O}_{j})\end{math}, and  \begin{math}\Hierar(i, j, \mathcal{O}_{j})\end{math} are countable orders.
\item Given a master finite order \begin{math}O_m\end{math},
  and a corresponding sequence of countable orders \begin{math}\mathcal{O}_{j}\end{math},
  \begin{math}\Domain(\sum(O_m, \mathcal{O}_j))\end{math} is the finite union 
  of the countable domains of the \begin{math}O_i\end{math}, hence is countable.
\end{itemize}
\end{demo}

\section{Orders that almost fit in our framework and open problems}
\label{section:orders_that_almost_fit_in_our_framework}

We start this section with the result of this article we like the most.
What about \rationals? How can rationals be efficiently TSO-encoded?
No solution seems to appear when we are only given the numerator and denominator,
since, when you compare two rationals, you multiply the numerator of one with the denominator of the other, and vice-versa,
 and you compare the two results of these multiplications.
But one can consider the continued fraction representing a rational:
\begin{math}\frac{p}{q} = n_0 + \frac{1}{n_1+\frac{1}{n_2+\frac{1}{n_3+\frac{1}{n_4+...}}}}\end{math}.
A number is rational, if and only if it can be represented by a finite continued fraction.
Assume we have non-negative rationals, since it is easy to add the sign with generalized sum.
Then for these rationals we have a sequence of positive integers.
When the integer of rank 0 is greater then the rational is greater, when the integer of rank 1 is greater then the rational is lower,
when the integer of rank 2 is greater then the rational is greater, when the integer of rank 3 is greater then the rational is lower, etc.
We have a problem for when this sequence finishes however since we have to apply lexicographic way of dealing with end of sequence alternatively with contre-lexicographic way of doing so.
We note here however that only \begin{math}n_0\end{math} may have value equal to zero (it is the case only when the rational is less than 1),
 since continued fractions integers are derived from gcd-algorithm,
 that the continued fraction may be written so that the last number is at least 2 if it is not \begin{math}n_0\end{math}.
It is thus easy to see by induction that any continued fraction without integer part is strictly between 0 and 1, and then:
\begin{itemize}
\item If we finished with a number of even rank, then the rational is lower than any other rational with the same beginning of continued fraction.
      We need to encode one more ``extended integer'' with positive infinite value because it will be compared with inverse integer order with the next integer.
\item If we finished with a number of odd rank, then the rational is greater than any other rational with the same beginning of continued fraction.
      Here again we need to encode one more ``extended integer'' with positive infinite value because it will be compared with integer order with the next integer.
\end{itemize}
Since positive integers are a suborder of the order \begin{math}O^{uint} = \Hierar(1, \omega, \mathcal{O}^{0,1})\end{math},
the order on extended integers is a suborder of \begin{math}O^{uint\_ext} = \Next(2, 3, (O^{0,1}, O^{uint}))\end{math}
(a leading 1 denotes positive infinity, a leading 0 denotes an integer).
We have that non-negative rationals order is a suborder of \begin{math}O^{\rationals^{+}} = \Next(1, \omega, ([O^{uint\_ext}, \Inv(O^{uint\_ext})]))\end{math}.

We let the reader write the expression for all rationals using generalized sum.
Note that it is our first use of an ultimately periodic sequence of period greater than 1.
Unfortunately it is not linear? time to compute the continued fraction of a rational number.
We may end-up with \begin{math}\Theta?(\MultiplyCost(\frac{m}{2n}) \times \log(\frac{m}{2n}) \times n)\end{math} complexity,
 if all rationals are roughly of size \begin{math}\frac{m}{n}\end{math}, 
with a denominator and a numerator of size approximately \begin{math}\frac{m}{2n}\end{math},
and \begin{math}\MultiplyCost(p)\end{math} denotes the time complexity of multiplication/division of integers
of \begin{math}p\end{math} bits
(gcd-algorithm to compute the continued fraction can be done with \begin{math}\log(p) = \log(\frac{m}{2n})\end{math}
such divisions).
Nevertheless, we have this theorem.

\begin{theorem}
\rationals\ can be almost efficiently embedded in a finite width tree structured order.
\end{theorem}

\begin{openproblem}
Prove or disprove that \rationals\ can be efficiently embedded in a  finite width tree structured orders.
(A linear time algorithm for computing the continued fraction of a rational would be a solution.)
\end{openproblem}

We now turn our attention on the many orders that do not directly fit in our framework.
We saw that \begin{math}\Next\end{math} had two simple linear extensions, \begin{math}\Lex, \ContreLex\end{math}.
But in fact, sequence termination could be ordered in between the elements of the next order in the sequence.
Let us call this kind of linear extensions middle-lexicographic, \begin{math}\MidLex\end{math};
for \begin{math}\mathcal{O}^{0,1}\end{math}, the choices of middle-lexicographic linear extensions are restricted to one choice:
sequence termination is between 0 and 1.
With \begin{math}\mathcal{O}^{0,1,2,...,9}\end{math}, we have many more choices, but still we could do something.
Indeed for a sequence of finite orders, we can embed each order in the order of cardinality one more;
this additional element would be the image of the sequence termination.
It would be very cumbersome with non-uniform sequence (sequence of period lengthier than one), but it would work.

With infinite orders, it may become very difficult.
For example, what can we do if we have a middle-lexicographic order on an uniform sequence of the rationals order
 (or more precisely of the order \begin{math}O^{\rationals}\end{math})
where we want to have sequence termination mapped to \begin{math}\pi\end{math}, or \begin{math}\sqrt{2}\end{math}?

\begin{openproblem}
Study in depth the frontiers of (finite width) tree structured orders.
\end{openproblem}

\begin{openproblem}
We have all these callback comparison functions in the wild.
Is it possible to code an algorithm that will take the assembly code, C code,
 or ``insert your favorite programming language here'' code of such a function
and extracts the corresponding finite width tree structured order definition/expression?
The problem must be undecidable but for ``common'' comparison functions it should be possible.
\end{openproblem}

Last but not least, we look at open problems around universality results.
Please read again the following definition, but this time thinking that the orders may be partial or total.
\begin{definition}[Universal order]
We say that an order \begin{math}O\end{math} is \emph{universal} for a class of orders \begin{math}\mathcal{A}\end{math},
 if for any order \begin{math}O' \in \mathcal{A}\end{math},
 there exists an embedding of \begin{math}O'\end{math} in \begin{math}O\end{math}.
An order embedding is an injective mapping such that order is preserved.
\end{definition}
``Order is preserved'' means that no order relation is removed or added (adding was not possible with total orders).
What we proved efficiently,
and that Cantor also proved with some minor addition (double each binary digit and put 0 on the last digit, 1 otherwise),
is that the partial order \begin{math}\Next\end{math} on a sequence of \begin{math}\omega\end{math} binary orders is universal for
the countable total orders.
Using the partial order \begin{math}\Next\end{math} limits comparison to solving the unique question if it exists.
We can say that all countable total orders have a representation that is ``questionable''.
The fact that this representation is ``questionable'' implies that this representation is well-ordered.
This is counter-intuitive if you think that rationals can be represented in a well-ordered way.
A ``questionable representation'' has two parameters: the maximum cardinality of the order-items it uses that we call its width,
and the ordinal length of the sequence of order-items.
(It reminds us of path-width and path-decompositions, but suited for orders.)
Countable total orders have a binary questionable representation of length \begin{math}\omega\end{math},
but they do not all have a unary questionable representation of length \begin{math}\omega\end{math}.
\begin{openproblem}
Is it true that any total order \begin{math}O\end{math} has a finite (binary?) width questionable representation?
(The length of the representation may adapt to the cardinality of \begin{math}O\end{math}.)
If not, what is the minimal cardinal of an order without finite (binary?) width questionable representation?
\end{openproblem}
Note that the same problem for partial orders has an immediate answer.
Indeed, the partial order \begin{math}POC\end{math} with four elements a, b, c, d such that a is less than c and b is less than d
does not admit a questionable representation.
(Two elements are not ordered if and only if they don't have a question if and only if one is the prefix of the other.
Thus, without loss of generality, a is a prefix of b.
But then, since c is more than a, it has a question with a, and since a is a prefix of b,
c has a question with b, and c is more than b, a contradiction.)
It seems we must consider partial questionable representations, where the order-items in the representation may be partial.
If the question yields two incomparable element-items, then the elements are incomparable.
Hence, we have the two following open problems.
\begin{openproblem}
Is it true that any partial order \begin{math}PO\end{math} has a total questionable representation
if and only if it does not contain some partial order \begin{math}POC\end{math}
taken in a finite set of forbidden finite partial orders?
\end{openproblem}
\begin{openproblem}
Is it true that any (countable) partial order \begin{math}PO\end{math}
has a (finite) partial questionable representation?
\end{openproblem}

\section{Tree Structured Orders Definition Language}
\label{section:TSODL}

We would like to create a language to create orders and sort software objects accordingly.
We hope the software and scientific community can contribute to this goal.
Here is an attempt at what it would look like.
It ressembles SQL, but with a different scope:
\begin{itemize}
\item it could be used for order by clauses of SQL queries or other query languages,
\item but it could also be used to generate code in your favorite programming language if you intend to sort objects without using a database,
\item or it could be used as an input of a library that provides functions to prepare the array to sort (TSO-encoding) and sort accordingly.
\end{itemize}
Both generated code or dynamic code in library could provide a comparison function and a ``nextification'' function,
computed from the tree structured order definition,
and both code could switch between comparison model and lexicographic model, whichever is faster,
according to the number of elements to sort and other parameters deduced from the tree structured order definition.

First each node of a tree structured order is centered around an object (or a structure in C language, etc.).
The root node is centered on the main objects, the ones that correspond to the wanted level of granularity.

Let us explain our ideas using an example from our current work in business software for freight forwarders.
Each night/morning, a freight forwarder receives trucks with freight from other freight forwarders.
For each truck and each day, we have an ``Arrival note'' in the database.
To this arrival note are linked shipments, each shipment has a certain number of handling units.
We have these three levels of granularity: arrival note (corse grained), shipments, handling units (fine grained).
Each of these three classes may have between a dozen and a few hundreds fields.
Assume we want to sort shipments according to the name of the freight forwarder that brought his truck (field found in the arrival note),
the weight of the freight in the truck by decreasing order (arrival note), the weight of the shipment by increasing order,
 the barcodes of the handling units.

A Tree Structured Order Definition for this sort would look like:
\begin{verbatim}
Next(
    current.arrival_note.forwarder_name STRING collation=UTF8_test,
    current.arrival_note.total_weight DOUBLE DESC,
    current.weight DOUBLE,
    Lex(
        1,
        0, //0 codes omega
        current.number_of_handling_units, //the actual prelude length
        current.array_of_handling_units, //the array of pointers to redefine current
        ([
          //current has been redefined for the suborders.
          current.barcode STRING collation=ASCII 
        ])
    )
)
\end{verbatim}

Note that we assumed denormalization or caching for the total weight of an arrival note.
With TSO-encoding, that value, if not directly available, would be computed once and stocked into the TSO-encoding.
With black box model and callbacks, you have to cache this value somehow, otherwise it will be computed on each comparison.

\section{Conclusion}

We did all this research because of the following reasons:
\begin{itemize}
\item linear time sorting was an old problem we were thinking about,
\item we encountered real life problems with MySQL optimizer where it would use an index instead of another because of the ORDER BY component of the query,
  and this choice was dramatic since the chosen index was not discriminating enough.
  Because of the number of rows, it would end up with an execution plan taking more than several minutes instead of an execution plan taking less than 3 seconds;
  When you have to optimize over and over dynamic queries with over 100 lines of SQL code with an ORDER BY that the user can choose among more than a dozen choices,
  and when you have almost no control on what does the optimiser,
  you have to add dynamicaly USE INDEX or FORCE INDEX to ensure that the execution plan is correct.
  It gets very frustrating that your DBMS does not have an option ``filtering is hard, sorting is easy''.
  We hoped to make sorting out of scope for index choice altogether.
  With SQL we noticed how \begin{math}\Lex, \Next\end{math} (partial) order was important.
\end{itemize}

We were thinking about optimizing DBMS all along, and more recently we thought it may also help search engines.
For search engines, we lack knowledge but probably the score of a webpage is coded into a float/double.
Since this score can be converted efficiently to unsigned int (see our library on GitHub),
and the number of webpages is enormous, it makes a good candidate for efficient use of our ideas.
Moreover all our results are straightforwardly parallelizable.

One of the interests of (efficient) universality results is that they open a new possibility:
Use optimized hardware for ``nextifying'' and sorting.
(We have specialized hardware for things as stupid as mining bitcoins;
it would be very interesting to have specialized hardware for sorting;
with comparison based sorting, because of the comparison function, you need a CPU,
so there is less room for a dedicated hardware.)

We benchmarked our ideas (see \url{https://github.com/LLyaudet/TSODLULS} and try it yourself) on our laptop with finite orders
((signed and unsigned) integers of 8, 16, 32, and 64 bits respectively, and floats and doubles),
and also with strings.
Although our ideas are memory expensive compared to comparison based sorting,
we succeeded to beat glibc comparison based qsort implementation in most cases
with finite orders.
We tested power of 2 number of elements between 1 and \begin{math}2^{25}\end{math}.
We were comparing our algorithm to glibc qsort in three settings:
\begin{itemize}
\item ``direct'' where the input is an array of keys (an element of a finite order),
\item ``in short cell'' where the element of the array is a cell containing a key that is an unsigned integer on 64 bits,
  and a pointer to the object to be sorted according to the key (128 bits for the short cell).
\item ``in long cell'' where the element of the array is a cell containing a pointer to the key, a pointer to the object to be sorted according to the key,
  the size taken by the key, and the size allocated for the key (it is 256 bits for a cell without the key).
\end{itemize}
Our sorting algorithm always works ``in cell'' and we added the cost of creating the ``in cell'' array and filling it in our benchmarks.
However, we implemented our algorithm in two variants: one for short cells and one for long cells.
The variant on short cells, when it can be used, is faster than the variant on long cells.
For finite orders, our algorithm on short cells is faster than glibc qsort, as soon as there are at least 64 elements to sort.
It can then be between 1.3 (random poor performance) and 14 times faster (on int8) but there is a lot of variations.
(Please launch the custom benchmark wrapper PHP script in TSODLULS with choice 1: qsort direct
and choice 2: TSODLULS\_sort\_radix8\_count\_insertion\_\_short with macraffs.
Your mileage may vary.)
It is fair to say that in our experiments, we were most of the time at least 2 times faster than glibc qsort;
and it was not uncommon to be 3 or 4 times faster.
We truly hope it may be \#ClimateChangeBrake.

Results for strings are not that good.
It seems that our algorithm on long cells is not cache efficient when dealing with long strings.
Hence, for a small number of elements we can be as bad as 40 times slower than qsort.
When the number of elements grows, it gets better, and we ended being 2 times faster with all optimizations activated.
(Please launch the custom benchmark for strings wrapper PHP script in TSODLULS with choice 1: qsort direct
and choice 2: TSODLULS\_sort\_radix8\_count\_insertion with macraffs.
The benchmark sorts strings of printable ascii characters with two settings:
raw order of the ascii characters, or custom collation to have the order AaBbCc...
Results are slower for custom collation, but they are ``less slower'' for our algorithm
than for glibc qsort, because we only apply the custom collation once during the ``nextification'',
whilst glibc qsort applies it for each comparison.
We tested with strings of length between 0 and 128 and no common prefix,
and with strings of length between 100 and 128 and a common prefix of length 90.
With a huge number of elements, we were still 2 or 5 times slower than qsort.
We finally succeeded to beat glibc for sorting long strings when the number of strings
is enough by optimizing the padding.
First padding was byte-level padding, hence we were multiplying the length of the strings by two,
because of the lexicographic ordering emulated by next partial order.
We saw that too much memory waste was the main problem and we implemented
bit-level padding because it was sufficient to increase the length of the strings by \begin{math}\frac{1}{8}\end{math}.
It was a huge improvement.
Your mileage may vary.)
If you look at the source code in detail,
you will probably agree that all the optimizations outside of the sorting algorithms
are too complicated for a simple task such as sorting.
It makes more important to design an order description language that will deal with this kind of code generation.

Micha\"el Rao suggested us to look at Boost C++ library.
It is clear there that the algorithmic ideas for sorting, and nextifying finite orders are not new.
We will try to improve our C library using ideas from Boost.

\acknowledgements
\label{section:acknowledgements}

We thank God: Father, Son, and Holy Spirit. We thank Maria.
They help us through our difficulties in life.

We thank Micha\"el Rao for his interest in our results and his comments.
We thank Pascal Koiran and Gilles Villard for useful corrections on the complexity of
results about rationals.

\nocite{*}
\bibliographystyle{abbrvnat}
% use the following instead if you encounter problems 
%\bibliographystyle{alpha}
\bibliography{LL2017OrdreLexico}
\label{section:bibliography}

\end{document}